   \documentstyle[12pt,ijmpe1,rotating,epsfig]{article}

\makeatletter

\def\maketitle{\par
 \begingroup
 \def\thefootnote{\fnsymbol{footnote}}
 \def\@makefnmark{\mbox{$^\@thefnmark$}}
 \@maketitle 
 \@thanks
 \endgroup
 \setcounter{footnote}{0}
 \let\maketitle\relax
 \let\@maketitle\relax
 \gdef\@thanks{}\gdef\@author{}\gdef\@title{}\let\thanks\relax}
\def\@maketitle{\vspace*{0.2cm} 
{\hsize\textwidth
 \linewidth\hsize \centering
 {\large \bf \@title \par} \vskip 0.3cm {\normalsize  \@author \par}}}

\def\thefootnote{\mbox{\noindent$\fnsymbol{footnote}$}}
\long\def\@makefntext#1{\noindent$^{\@thefnmark}$#1}
\makeatother

\begin{document}
\begin{flushright}
TRI--PP--99--28\\
Sept 1999
\end{flushright}

\title{NUCLEON-NUCLEON PARITY VIOLATION EXPERIMENTS\footnote{
Work supported in part by the Natural Sciences and Engineering Research
    Council of Canada}}
\author{WILLEM T.H. VAN~OERS\footnote{For the E497 Collaboration: J.~Birchall, 
J.~Bland, J.D.~Bowman,
    C.A.~Davis, P.W.~Green, A.A.~Hamian, R.~Helmer, S.~Kadantsev, Y.~Kuznetsov,
    R.~Laxdal, L.~Lee, C.D.P.~Levy, R.E.~Mischke, S.A.~Page, W.D.~Ramsay,
    S.D.~Reitzner, G.~Roy, G.M.~Stinson, V.~Sum, N.A.~Titov, 
    W.T.H.~van~Oers, R-J.~Woo, A.N.~Zelenski.}\\
Institut f\"ur Kernphysik, Forschungszentrum J\"ulich, \\
D-52425 J\"ulich, Germany\\
and\\
Department of Physics, University of Manitoba, \\
Winnipeg, MB, Canada R3T 2N2\\ 
and \\
TRIUMF, 4004 Wesbrook Mall, Vancouver, B.C., \\
Canada V6T 2A3}

\maketitle

\begin{abstract}
Measurements of parity-violating longitudinal analyzing powers $A_z$              
(normalized asymmetries) in polarized proton-proton scattering and in polarized 
neutron capture on the proton $(n - p \to d - \gamma)$ provide a unique window on  
the interplay between the weak and strong interactions between and within 
hadrons. Several new proton-proton parity violation experiments are presently   
either being performed or are being prepared for execution in the near future:  
at TRIUMF at 221 MeV and 450 MeV and at COSY (Forschungszentrum J\"ulich) in the
multi-GeV range. A new measurement of the parity-violating $\gamma$ ray 
asymmetry with a ten-fold improvement in the accuracy over previous
measurements is being developed at LANSCE. These experiments are intended to
provide stringent constraints on the set of six effective weak meson-nucleon
coupling constants, which characterize the weak interaction between hadrons
in the energy domain where meson exchange models provide an appropriate
description. The 221 MeV $p - p$ experiment is unique in that it selects a 
single transition amplitude (${^3P_2} - {^1D_2}$) and consequently constrains the
weak meson-nucleon coupling constant $h_\rho^{pp}$. The $n - p \to d - 
\gamma$
experiment is mainly sensitive to the weak pion-nucleon coupling constant
 $f_\pi$. Together with the existing $p - p$ parity violation experimental results
one may be able to delineate the various weak meson-nucleon coupling 
constants. The TRIUMF 221 MeV $p - p$ parity violation experiment will be
described in some detail. Other parity violation nucleon-nucleon and nucleon-
very-light-nucleus experiments are commented on.
The anomalous result obtained at 6 GeV/c on a water target requires that a 
new multi-GeV $p - p$ parity violation experiment be performed.
\end{abstract}
\begin{center}
(paper submitted to International Journal of Modern Physics E)
\end{center}

\section{Introduction}
                     
   Because flavour changing neutral currents are almost completely suppressed
by the G.I.M. mechanism, the study of hadronic neutral currents in nuclear
systems provides a unique window on weak neutral currents. Parity violation in
nuclear systems is the only flavour conserving process in which hadronic weak
neutral currents can be observed. Observations of parity violation in the
nucleon-nucleon $(N-N)$ systems are complementary to studies of
parity violation in electron-proton scattering, the next generation of which
is focussing on constraining the possible contributions of strange quarks from
the sea to the nucleon form factors and the quest for "new physics".

   At low and intermediate-energies, the parity violating weak $N-N$ interaction
can be described in terms of a meson exchange model involving a strong
interaction vertex and a weak interaction vertex (assuming one-boson
exchanges). The strong interaction vertex is generally well understood; it
is represented by the conventional meson-exchange parameterization of the
nucleon-nucleon interaction. The weak interaction vertex is calculated from
the Weinberg-Salam model assuming that the $W$- and $Z$-bosons are
exchanged between the intermediate mesons ($\pi$, $\rho$, and $\omega$) and
constituent quarks of the nucleon. The parity violating interaction can
then be described in terms of seven weak meson-nucleon coupling constants.
The six weak meson-nucleon coupling constants ($f_\pi$, $h^0_\rho$,
$h^1_\rho$, $h^2_\rho$, $h^0_\omega$, $h^1_\omega$, with the subscripts
indicating the exchanged meson and superscripts indicating isospin changes)
have been calculated by Desplanques, Donoghue, and Holstein (DDH)$^1$,
synthesizing the quark model and SU(6) and treating strong interaction
effects in renormalization group theory. The seventh weak meson-nucleon
coupling constant $h^{'1}_\rho$ is estimated to be smaller and is often
deleted from further consideration. DDH tabulated ``best guess values" and
``reasonable ranges" for the six weak meson-nucleon coupling
constants. Similar calculations have been made by Dubovik and Zenkin
(DZ)$^2$. Extending the earlier work in the nucleon sector, Feldman,
Crawford, Dubach, and Holstein (FCDH)$^3$ included the weak
$\Delta$-nucleon-meson and weak $\Delta$-$\Delta$-meson parity
violating vertices for $\pi$, $\rho$, and $\omega $ mesons. The latter
authors also present ``best guess values" and ``reasonable ranges" for the
six weak meson-nucleon coupling constants. Using the expressions of an
earlier paper by Desplanques (D)$^4$ FCDH present a third set of weak
meson-nucleon coupling constants. It is apparent that these coupling
constants carry considerable ranges of uncertainty (see Table 1). Taking
into account the more recent nuclear parity violation experiments,
Desplanques$^5$ argues for a reduced value and range for the weak
meson-nucleon coupling constant $f_\pi$. The weak meson-nucleon coupling
constants have also been calculated by Kaiser and Meissner (KM)$^6$ within
the framework of a nonlinear chiral effective Lagrangian which includes
$\pi$, $\rho$, and $\omega$ mesons. In this model $ f_\pi$ is considerably
smaller than the ``best guess value" of DDH or FCDH (Table 1).
Furthermore, a non-zero and non-negligible value for the seventh weak
meson-nucleon coupling constant $ h_\rho^{'1}$ was found. The parity
violating $\pi \Delta N$ vertex plays an important role in elastic and
inelastic proton-proton scattering above the pion production threshold. The
latter authors$^7$ find that the isoscalar parity violating $\pi \Delta N$
coupling constant $h^0_{\pi \Delta N}$ vanishes identically, but that the
isovector parity violating coupling constant $ h^1_{\pi \Delta N}$ has a
strength of 2.1$\times$10$^{-8}$ and that the parity violating $\rho \Delta
N$ and $\omega \Delta N$ couplings generally cannot be neglected. Meissner
and Wiegel$^8$ find a sizeable enhancement of $ f_\pi$ in the framework of
a three flavor Skyrme model compared to previous calculations in two flavor
models $(f_\pi$ ranges from 0.8 to 1.3 $\times$ 10$^{-7}$). Holstein$^9$
has re-examined the quark model calculations and has concluded that a small
value of $ f_\pi$ cannot be understood unless the current algebra quark
mass values are increased by about a factor of two over the original
Weinberg values, which tends to produce a similar suppression of
theoretical estimates in other areas, e.g., the $\Delta${I} = 1/2 rule. It
appears that the theoretical situation regarding $ f_\pi$ is not very well
settled. For a recent review see Haeberli and Holstein.$^{10}$

   A complete determination of the six weak meson-nucleon coupling constants
demands at least six experimental, linearly independent combinations of the
weak meson-nucleon coupling constants. But as of to date there do not exist
enough experimental constraints of the required statistical significance. This 
situation can only be remedied by performing a set of judiciously chosen,
precision parity violation experiments.

   Impressively precise measurements of the proton-proton parity violating
longitudinal analyzing power have been made at 13.6 MeV
[$A_z $ = (-0.93 $\pm $ 0.20 $\pm $ 0.05) $\times$ 10$^{-7}$] at the University of Bonn$^{11}$ and
at 45 MeV [$A_z$ = (-1.57 $\pm$ 0.23) $\times$  10$^{-7}$] at the Paul Scherrer Institute
(PSI)$^{12}$. Here $A_z$ is defined as $A_z = (\sigma^+ -
\sigma^-)/(\sigma^+  + \sigma^-)$, where $\sigma^+$ and $\sigma^-$ represent the scattering cross
sections for polarized incident protons of positive and negative helicity,
respectively, integrated over a range of angles determined by the acceptance
of the experimental apparatus in question. A non-zero value of $A_z $ implies
parity violation due to the non-zero pseudo-scalar observable
$\vec{\sigma}$.$\vec{p}$ with $\vec{\sigma}$ the spin and $\vec{p}$ the momentum of
the incident proton. From the PSI measurement at 45 MeV and the 
$\sqrt{E}$ energy
dependence of $A_z$ at lower energies, one can extra\-polate $A_z $ at 13.6 MeV to be
$A_z $ = (-0.86 $\pm $ 0.13) $\times$ 10$^{-7}$. There exists thus excellent agreement 
between the above two lower energy measurements. Both results allow determining
a combination of the effective $\rho$ and $\omega$ weak meson-nucleon coupling
constants $A_z$ = 0.153$h^{pp}_\rho$ $+$ 0.113h$^{pp}_\omega$, with
 $ h^{pp}_\rho = h^0_\rho + h^1_\rho + h^2_\rho/\sqrt{6}$ 
and $ h^{pp}_\omega = h^0_\omega + h^1_\omega$. It should be noted that a
measurement of $A_z$ in $p-p$ scattering is sensitive only to the short
range part of the parity violating interaction (parity violating $\pi^0$
exchange would simultaneously imply CP violation and is therefore
suppressed). However, above the pion production threshold, where
intermediate $\Delta $ states become important, parity violating charged
pion-exchanges play a non-neglegible role.  In fact Silbar, Kloet,
Kisslinger, and Dubach$^{13}$ find that the pion-exchange contribution to
$A_z$ has both inelastic and elastic scattering contribution components and
is sizable even below the pion production threshold. These authors also
argue that strong distortions enhance the magnitude of this contribution.

   With the $p-p$ strong interaction phases known from phase shift analyses and
given a set of the parity violating mixing angles, one can calculate both the
angular and energy dependence of $A_z$.$^{14}$ A partial wave decomposition
allows the various contributions to $A_z$ to be separated out. The mixing
angles are directly related to the parity violating transition amplitudes
(${^1S_0}$ - ${^3P_0}$), (${^3P_2}$ - ${^1D_2}$), (${^1D_2}$ - ${^3F_2}$),
 (${^3F_4}$ - ${^1G_4}$), etc.
In practise it is almost impossible to measure the angular dependence of the
longitudinal analyzing power $A_z$ and consequently one has to resort to
measuring an angle-averaged longitudinal analyzing power $A_z$ while at the
same time excluding the forward, Coulomb scattering dominated, angular region.
A determination of $A_z$ can be accomplished through a scattering measurement,
restricted to the low-energy region, or through a transmission (attenuation)
measurement at intermediate and higher energies.
The energy dependence of the first two parity violating transition
amplitudes contributing to $A_z$ is shown in Fig.~1.$^{15}$ For energies
below 100 MeV essentially only the first parity violating transition
amplitude (${^1S_0}$ - ${^3P_0}$) contributes. One notices the increase in importance
of the second parity violating transition amplitude for energies above
100 MeV. The theoretical (${^1S_0}$ - ${^3P_0}$) contribution to $A_z$ was normalized
to the experimental datum at 45 MeV.$^{15}$. The contribution of the next higher
order (third) transition amplitude (${^1D_2}$ - ${^3F_2}$) is negligibly small.

   There exists a further $p-p$ parity violation measurement at 800 MeV with
$A_z = (2.4 \pm 1.1)  \times 10^{-7}$.$^{16}$ Interpretation of the latter result in
terms of the effective $\rho$ and $\omega$ weak meson-nucleon coupling constants
is more difficult due to the presence of a large inelasticity
(pion production). 

   Other $N-N$ measurements have dealt with the circular polarization of the
gamma-rays in $n-p$ capture$^{17}$ or with the longitudinal analyzing power in
$n-p$ capture with polarized incident cold neutrons.$^{18}$ For low-energy
neutrons parity violation in the reaction $n-p \to d-\gamma $ is almost entirely
due to weak pion exchange, as calculated by  Adelberger and Haxton$^{19}$ and
corroborated 
in an earlier
calculation of Desplanques and Missimer.$^{20}$ The final result obtained in the
last measurement is $A_\gamma $ = $(-1.5 \pm 4.7) \times 10^{-8}$.$^{18}$ A new
measurement of the parity violating longitudinal analyzing power $A_\gamma$ is
being prepared at the Los Alamos Neutron Science Center (LANSCE), aiming at a
ten-fold improvement in accuracy (to a precision of $\pm 0.5 \times 10^{-8}$, which
will determine $f_\pi$ to $\pm 0.4 \times 10^{-7}$).$^{21}$ In the 
experiment, neutrons
from the spallation source are moderated by a liquid hydrogen moderator. With
the spallation source pulsed, the neutron energy can be determined through
time-of-flight measurements. The cold neutrons are polarized by transmission
through polarized $^{3}$He gas; the neutron spin direction can be subsequently
reversed by a RF  resonance spin flipper. The neutrons are then guided to a
liquid para-hydrogen target which is surrounded by an array of gamma-ray
detectors. The parity violating helicity dependence in the inverse reaction,
the photodisintegration of the deuteron by circularly polarized gamma-rays,
was measured to be $(2.7 \pm 2.8)  \times 10^{-6}$ for Bremsstrahlung with an
endpoint of 4.1 MeV and ($7.7 \pm 5.3) \times 10^{-6}$ for an endpoint of 3.1 MeV,
essentially a null result.$^{22}$ A new parity violation experiment, measuring
the helicity dependence of the photodisintegration of the deuteron with
circularly polarized gamma-rays, may be possible at the HIGS facility
of the Triangle Universities Nuclear Laboratory. The very intense gamma-ray
beams are obtained from Compton back-scattering using free electron lasers.

   The higher flux of neutrons, which will become available from the new cold
neutron source at ILL, should make it possible to perform a new generation of
$n-p$ parity violation experiments with greatly improved precision. A
measurement of the neutron spin rotation in parahydrogen could accomplish the
same objective as a measurement of $A_\gamma$, in that both observables depend
almost entirely on the weak pion-nucleon coupling constant. The expected
effect is of the order $10^{-6}$ rad/m. A University of Washington based group
has proposed such an experiment.$^{23}$ and expects to achieve a sensitivity of
$4\sigma$ in 30 days of data taking. A measurement of the parity violating
spin rotation of cold neutrons through liquid helium is ongoing at NIST.$^{24}$
This rotation, $\Phi (\vec{n},\alpha)$, is sensitive to the weak pion-nucleon
coupling constant, and in conjunction with the result for the $ p-\alpha$
longitudinal analyzing power, $A_z(\vec{p}, \alpha)$, obtained at 45 MeV at
PSI$^{25}$ will determine $ f_\pi$. A first result of $\Phi(\vec{n},\alpha)$
equals (8.0 $\pm $ 14 [stat] $\pm $ 2.2 [syst]) $\times $ 10$^{-7}$ rad/m. With
$A_z(\vec{p},\alpha)$ = -(3.34 $\pm $ 0.93) $\times$ $10^{-7}$ and the 
DDH ``best values"
for $h^0_\rho, h^1_\rho, h^0_\omega, h^1_\omega $ one deduced that
 $ f_\pi = -(1.75 \pm 10.5) \times 10^{-7}$. The accuracies obtained to date in parity
violation measurements in the $n-p$ system do not suffice to constrain in a
significant way the weak meson-nucleon couplings. For a plot of the
constraints on the isoscalar and isovector weak meson-nucleon coupling
constants see Fig. 2 (see also Ref.~24).

   Following the approach of Adelberger and Haxton$^{19}$, one can fit the more
significant nuclear parity violation data, using theoretical constraints, by
the two parameters $f_\pi $ and $(h^0_\rho + 0.6 h^0_\omega)$. This leaves the
experimental value of $f_\pi = (0.28^{+0.89}_{-0.28}) \times 10^{-7}$, as deduced
from the circular polarization of the 1.081 MeV $\gamma$-rays from 
$^{18}$F
(Ref.26), at the border of the deduced range so determined.$^{10}$ See Table 1,
last two columns.

   The definitive non-zero result obtained by Wood et al.$^{27}$ in a
measurement of the anapole moment of $^{133}$Cs has been analyzed by Flambaum
and Murray$^{28}$ to extract a value for $f_\pi$. The resulting value of
 $f_\pi $ = $(9.5 \pm 2.1$[exp.] $\pm $ 3.5[theor.]) $\times$ $10^{-7}$ is a factor of two 
larger than the DDH ``best guess value" (see Table 1) and a factor of seven
larger than the upper limit set by the $^{18}$F results. But Wilburn and
Bowman$^{29}$ argue that the anapole moment is sensitive to a combination of
 $ f_\pi $ and $ h^0_\rho~(f_\pi + 0.69 h^0_\rho)$ and therefore to deduce
 $ f_\pi $ one must know $ h^0_\rho$, which still carries a considerable range of
uncertainty. Furthermore, the result from $^{133}$Cs is inconsistent with an
earlier null measurement of the anapole moment of $^{205}$Tl.$^{30}$ One can then
ask the question if there exists a dependence of the weak meson-nucleon
coupling constants on the nuclear medium.

   Fig.1 shows that there exists a unique feature at an energy of about
230 MeV: the (${^1S_0}$ - ${^3P_0}$) transition amplitude contribution integrates to
zero. This reflects a change in sign of both the ${^1S_0}$ and 
${^3P_0}$ strong
interaction phases near 230 MeV and is completely independent of the weak
meson-nucleon coupling constants. The absolute scale and sign of the ordinate
in Fig.~1, $(A_z)$, are determined by the weak interaction. Neglecting a small
contribution ($\approx$5\%) from the (${^1D_2}$ - ${^3F_2}$) transition amplitude, a 
measurement of $A_z $ near 230 MeV constitutes a measurement of the contribution
of the (${^3P_2}$ - ${^1D_2}$) transition amplitude. Simonius$^{31}$ has shown that the
 (${^3P_2}$ - ${^1D_2}$) transition amplitude depends only weakly on $\omega$-exchange
(to an extent determined by the choice of the strong vector meson-nucleon
coupling constants from various ${N-N}$ potential models), whereas $\rho$-exchange
and $\omega$-exchange contribute to the (${^1S_0}$ - ${^3P_0}$) transition amplitude
with equal weight. Consequently, a measurement of $A_z $ at an energy of 230 MeV 
presents a determination of $ h^{pp}_\rho$. The energy dependence of the real
parts of the $p-p$ phase shifts predicts that, neglecting inelasticity, the
(${^1D_2}$ - ${^3F_2}$) transition amplitude changes sign at about 650 MeV, and that
the (${^3P_2}$ - ${^1D_2}$) transition amplitude changes sign at about 950 MeV.
Precision measurements of $A_z$ at both 230 MeV and 650 MeV could provide
another determination of both $h_\rho^{pp}$ and $h_{\omega}^{pp}$ (in addition
to a combination of the low energy measurements with a measurement at 230 
MeV).

   Various theoretical predictions of the $p-p$ longitudinal analyzing power $A_z$,
based on meson-exchange models, have been reported: at 230 MeV the values of
$A_z$ are = $+0.7 \times 10^{-7}$ (Ref.~32), $+0.6 \times 10^{-7}$ (Ref.~15), 
and $+0.4 \times 10^{-7}$
(Ref.~33). Extensions to the one-boson exchange model have been made to
include $\pi-\pi $ and $\pi-\rho $ exchanges via $N-\Delta $ and $\Delta-\Delta$
intermediate states to which the (${^3P_2}$ - ${^1D_2}$) transition amplitude is
particularly sensitive.$^{13,32}$ For instance, Iqbal and Niskanen$^{32}$ find that
the $\Delta $ isobar contribution at 230 MeV (which is dependent on $f_\pi)$ may
be as large as the $\rho$-exchange contribution, enhancing the value of $A_z$ by
a factor of two. What is required is a self-consistent theoretical calculation
of $A_z$, avoiding possible double counting and taking into account that the
value of $f_\pi$ is constrained by experiment to be rather small. The latter
assumes that the discrepancy which has arisen by the measurement of a large
anapole moment of $^{133}$Cs has been resolved. It is to be noted that the 
prediction of $A_z $ by Nessi-Tedaldi and Simonius$^{15}$ had been normalized to
the experimental datum at 45 MeV as remarked above. Driscoll and Miller$^{33}$
predict a rather small value for $A_z$; their theoretical curve does not agree
very well with the two low-energy data. Scaling to the low-energy experimental
data at 13.6 and 45 MeV, would give an even smaller value for $A_z $ at 230 MeV
indeed. Considering all of the above, a measurement of $A_z $ at 230 MeV to an
accuracy of $\pm 2 \times 10^{-8}$ would provide a most important determination of
parity violation in $p-p$ scattering. Following the formalism of Simonius$^{15}$
and taking into account the finite geometry of the TRIUMF $p-p$ parity
violation experiment described below one can show that
 $A_z$(221 MeV) = -0.0296 $\times$ $h^{pp}_\rho$.

\section{The TRIUMF 221.3 MeV $p-p$ Parity Violation Experiment (see 
Fig.~3)}

   In the current TRIUMF experiment a 200 nA proton beam with a polarization
of 0.80 is incident on the 0.40 m long LH$_2$ target, after extraction from the
optically pumped polarized ion source (OPPIS), passing a Wien filter in the
injection line, acceleration through the cyclotron to an energy of 221.3
MeV, and multiturn extraction of the H$^-$ polarized ions. The incident energy
was chosen to correspond to integration to zero of the contribution of the
(${^1S_0}$ - ${^3P_0}$) transition amplitude over the finite acceptance of the parity
violation measuring apparatus. A combination of 
solenoid-dipole-solenoid-dipole magnets on the external beam line provides a
longitudinally polarized beam with either positive or negative helicity. The
longitudinal analyzing power $A_z$ follows from the helicity dependence of the
$p-p$ total cross section as determined in precise measurements of the
normalized transmission asymmetry through the 0.40 m long LH$_2$ target:
$A_z = -(1/P)(T/S)(T^+ - T^-)/(T^+ + T^-)$, where $P$ is the incident beam
longitudinal polarization, $T = 1 - S$ is the average transmission through the
target, and the $+$ and $-$ signs indicate the helicity state.
   
   There are many other effects that can cause such a helicity correlated
change in transmission. Very strict constraints are imposed on the incident
longitudinally polarized beam in terms of intensity, transverse x (horizontal)
and y (vertical) beam position and direction, beam width (given by $\sigma_x$
and $\sigma_y$), longitudinal polarization $(P_z)$, transverse polarization (given
by $P_y$ and $P_x$), first moments of the transverse polarization (given by
$<xP_y>$ and $<yP_x>$), and energy, together with deviations of the transmission
measuring apparatus from spatial symmetry. Helicity correlated modulations in
the beam parameters originate at OPPIS, but can be amplified by the beam
transport through the injection beam line, the cyclotron accelerator and the
extraction beam line. Residual systematic errors, arising from the imperfections of
the incident beam and the response of the transmission measuring apparatus,
are individually not to exceed one tenth of the expected value of $A_z$
(or $6 \times  10^{-9}$). Particular troublesome are the first moments of the residual
transverse polarization (so called ``circulating" polarization profiles),
as well as energy changes. In addition to the strict constraints imposed on
the incident beam parameters and on the quality of the measuring apparatus,
the approach which is being followed is to further measure the sensitivity
or response to residual imperfections, to monitor these imperfections during
data taking, and to make corrections where necessary. Random changes of the
incident beam parameters cause a dilution of the effect to be measured and
therefore longer data taking times in order to arrive at the desired
statistical error.

   Helicity changes are implemented through shifts in the linearly polarized
laser light frequency (a frequency change of 94 GHz at a magnetic field of
2.5~T),
 minimizing helicity correlated changes in the  
accelerated beam parameters. The optimum beam current at the parity
violation measuring apparatus is 200 nA; to achieve very small helicity
correlated modulations, most of the OPPIS intensity is sacrificed for beam
quality. As an example, the RF bunchers in the injection beam line, used to
enhance the cyclotron transmission by a factor of four, also increase the
sensitivity to energy modulations by more than two orders of magnitude and
therefore cannot be used in the parity violation experiment. A high
brightness OPPIS was developed for the parity violation experiment;
simultaneous high-current OPPIS development$^{35}$ for high-energy
accelerators has contributed greatly to the parity violation
experiment. The Faraday effect provides a means to monitor and control
on-line 
the polarization of the Rubidium vapour (which produces electron
polarized $\vec{H}^0$ through charge exchange) by using a probe laser. The
polarization of the linearly polarized laser light is rotated by an angle
proportional to the Rubidium vapour polarization. The polarizations of the
Rubidium vapour in the two helicity states are maintained to be the same
within 0.005 close to 1.00. The Faraday rotation measurement also provides
confirmation of the helicity state at OPPIS.
   
   To aid in tuning the extracted polarized proton beam, various retractable
horizontal and vertical wire chambers are placed along the beam line. 
Following the second dipole magnet, where the polarization direction has both
a longitudinal and a horizontal sideways component, a four branch polarimeter
measures the transverse polarization components, while a beam energy monitor
measures the relative energy of the proton beam with a precision of
$\pm $ 20 keV during a one hour data taking run. Measurements are made several
times each data taking period. The absolute energy has to agree within a
few MeV with the energy for which the (${^1S_0}$ - ${^3P_0}$) transition contribution
integrates to zero, taking into account the finite geometry of the
transmission measuring apparatus; but note that any changes in energy greater
than 40 keV will introduce transverse polarization components at the
transmission measuring apparatus in excess of 0.001 for the canonical setting
of all beam transport magnetic elements. All beam transport magnetic elements
have their excitations monitored on a continuous basis (super-conducting
solenoids - currents; dipole magnets using NMR probes; quadrupole magnets
using Hall probes).

   Figure 4 gives a three dimensional view of the downstream part of the
experimental setup. The longitudinally polarized beam, incident from the
lower right, passes first a series of diagnostic devices - a set of three
beam intensity profile monitors (IPMs), and a pair of transverse polarization
profile monitors (PPMs) - before reaching the LH$_2$ target which is preceded
and followed by transverse electric field ionization chambers (TRICs) to
measure the beam current. Note the position of the third IPM.

   Ideally the beam transport is to produce an achromatic waist downstream of
the LH$_2$ target halfway into the second transverse electric field ionization
chamber (TRIC-2). The beam is converging downstream of the last quadrupole
magnet triplet; its dimensions at the location of the first two IPMs are
characterized by ($\sigma_x = \sigma_y$ and the x and $\theta$ parameters decoupled 
from the y and $\phi$ parameters).
Helicity changes can also be accomplished by reversing the currents of the
two super-conducting solenoids. The rotation of phase space introduced by the
super-conducting solenoids is negated by rotation around the beam line axis
of various sets of quadrupole magnets. Consequently the reversal of
the solenoid currents requires simultaneous reversal of the quadrupole
rotation angles. Empirical beam line tunes were developed which are very
similar for the `normal' and `reversed' solenoid excitation settings. These
empirical tunes deviated in certain aspects from the specified, calculated,
beam line tunes.  

   Helicity correlated current modulations, expressed as $\Delta{I}/I =
(I^+ - I^-)/(I^+ + I^-)$, introduce a systematic error $\Delta{A_z}$ through
non-linearity of the TRICs and associated electronics. The parity detection
apparatus attains minimal sensitivity to these current modulations by
precision analog subtraction of the current signals of the two TRICs.
Controlled helicity correlated current modulations are needed for tuning
the precision subtractor circuitry for minimal sensitivity. These control
measurements are provided by an auxiliary argon-ion laser beam which co-propagates
with the H$^-$ beam along the 30 m long horizontal section of the injection
beam line (as shown in Fig.~3), neutralizing through photodetachment a fraction
of the H$^-$ ions along its path. The photodetachment laser is interrupted
synchronously with the parity spin sequence, so that the beam current in every
second spin `off' data taking cycle is modulated at the 0.1\% level, giving
the desired control measurements. Improvements to the stabilization of the
Electron Cyclotron Resonance (ECR) primary proton source, the Rubidium vapour
thickness in the polarizer cell, OPPIS high voltage levels, the injection beam
line elements (in particular the Wien filter), and the cyclotron accelerator
have led to a reduction in the width of $\Delta${I}/I to about the required
 $1\times 10^{-5}$. Table 2 presents a comparison of calculated and measured
sensitivities to systematic errors and the precision required in measuring
the various parameters.

   The IPMs$^{36}$, which are based on secondary electron emission from thin, 
  3~$\mu$m thick nickel foil strips placed between 8~$\mu$m aluminium high 
 voltage foils, measure the beam intensity profile with harps of 31 
 strips (1.5 mm wide, separated 2.00 mm center to center) in both the 
 vertical (x-profile) and horizontal (y-profile) directions.  (Fig.~5) 
 The third IPM placed just in front of the LH$_2$ target has 10~$\mu$m 
 thick nickel foil strips (2.5 mm wide by 3.00 mm center to center).  
 The harp signals are individually amplified and digitized to provide 
 the beam intensity profiles in x and y, from which the beam positions 
 are derived.  Analog beam centroid evaluators (BCEs) in turn determine 
 the beam intensity profile centroids at two locations through 
 appropriate integration of the discrete distributions; a corresponding 
 normalized error signal is used to drive feedback loops to a pair of x 
 and y fast, ferrite-cored steering magnets.  This allows the beam 
 intensity profile centroids to be kept fixed within 1~$\mu$m with an 
 offset less than 50~$\mu$m from the `neutral' axis in both x and y 
 during a one hour data taking run.  The beam intensity profile widths 
 maintained during data taking are: IPM-1 $\sigma_x$ = $\sigma_y$ = 5 
 mm; IPM-2 = 4 mm; IPM-3 = 6 mm.  Typical values for the modulations in 
 position and width in a one hour data taking run are 
 $\Delta$x, $\Delta$y $<$ (0.5 $\pm$ 0.3) $\mu$m and $\Delta \sigma_x$, 
 ${\Delta \sigma}_y$ $<$ (1.0 $\pm$ 0.6) $\mu$m.  Sensitivities to helicity 
 correlated position and size modulations are determined with enhanced 
 modulations introduced using the fast, ferrite-cored magnets 
 synchronized to the helicity sequence of the experiment to allow for 
 off-line corrections.  The advantage of the BCE based position 
 stabilization system over the earlier median intensity based position 
 stabilization system (which used the amplified signals from split foil 
 monitors) is a reduction in sensitivity to shape fluctuations in the 
 beam intensity profiles.
      
   The PPMs are based on $p-p$ scattering using CH$_2$ targets. Scattered protons
are detected in a forward arm at 17.5$^\circ$ with respect to the incident beam
direction with a pair of scintillation counters. The solid angle defining
scintillator is rotated around an axis perpendicular to the scattering plane
to compensate for changes in solid angle and differential cross section,
when the CH$_2$ target blade moves through the beam (see Fig.~6). Coincident
recoil protons are detected in a backward arm at 70.6$^\circ$ with respect to the
incident beam direction with a recoil scintillation counter; a second
scintillation counter acts as a veto for higher energy protons from
$^{12}$C(p,p)X. Each PPM contains detector assemblies for `left' scattered
protons, `right' scattered protons, `down' scattered protons, and `up'
scattered protons. The targets consist of CH$_2$ blades, 1.6 mm wide by 5.0 mm
thick along the incident beam direction. The blades move through the beam on
a circle of 0.215 m at a frequency of 5 Hz. Each PPM has four blades; two
which scan the polarization profile in the horizontal direction and allow for
determining the quantity $(L-R)/(L+R)$ and therefore $P_y$ as a function of x
for each of the two helicity states, and two which scan the polarization
profile in the vertical direction and allow for determining the quantity
$(D-U)/(D+U)$ and therefore $P_x $ as a function of y for each of the two
helicity states. Residual transverse polarizations (which change sign with
helicity reversals) can cause a false $A_z $ via the parity allowed transverse
analyzing power, which produces asymmetric scattering in the LH$_2$ target.
The sensitivities to transverse polarizations are dependent on the
incident beam position and on the geometry of the parity violation measuring
apparatus. Proper beam tuning greatly reduces (by a factor of approximately
20) the transverse polarizations at the LH$_2$ target compared to those at
injection of the cyclotron. Both the sensitivities and the `neutral axis'
are determined by introducing enhanced transverse polarizations $P_x$ and $P_y$.
Non-zero first moments of the transverse polarizations, $<xP_y>$ and 
$<yP_x>$,
can arise from an inhomogeneous polarization distribution of the cyclotron
beam at the stripper foil location, and from spin precession in the magnetic
field gradients at the entrance and exit of the solenoids, dipole magnets,
and quadrupole magnets. The latter combined have been estimated to contribute
less than 10 $\mu$m to the transverse polarization moments. The sensitivity
to intrinsic first moments is determined by deducing the correlations between
apparent $A_z$ and the $<xP_y>$ and $<yP_x>$ as measured by the PPMs. The moments
of transverse polarization exhibit a random variation from run to run and
reach values as high as 30 $\mu$m as measured in one hour data taking runs.
In a drift space first moments vary linearly with position along the beam
line, permitting the adjustments of the beam transport parameters such that
the first moments pass through zero at a point which minimizes their effect.
Reducing the first moments of transverse polarization was one of the more
challenging aspects of the experiment; it required frequent retuning of
OPPIS, the cyclotron, and beam transport systems. With measured
sensitivities to $<xP_y>$ and $<yP_x>$ of $+5 \times 10^{-5}$ mm$^{-1}$ 
and $-8 \times 10^{-5}$
mm$^{-1}$, respectively, one needs to measure these to a precision of
$12 \times 10^{-5}$ mm and $8 \times 10^{-5}$ mm, respectively, over the course of the
experiment in order that their possible individual contribution to the error
in $A_z $ is less than $6 \times 10^{-9}$. 
 
   With two PPMs, each with four blades, the spin flip or helicity flip rate
becomes 40 Hz, i.e., in one cycle all eight blades of the two synchronized
PPMs (four blades of PPM-1 and four blades of PPM-2) will pass once
through the beam. The master clock for sequencing the entire experiment,
including helicity changes, is derived from the PPMs shaft encoders. In order
to suppress up to second order other than helicity correlated effects
stemming from changes in the incident polarized beam parameters, a cycle
consists of the following sequence of helicity states: $+- - + - + + -$,
lasting 200 ms. Eight cycles are repeated, with the first helicity state
chosen so as to form an eight by eight symmetric matrix of helicity states
(designated a super-cycle). After three of such super-cycles, the pumping laser
light is blocked by a shutter during one super-cycle for control measurements.
In every other super-cycle used for control measurements, the sensitivity to
helicity correlated changes in the incident beam intensity is measured. Of
each helicity state of 25 ms duration, a little more than 1 ms is reserved
for the polarization to stabilize following a helicity change, the next
6.4 ms is used for the measurement of the transverse polarization (one of the
CH$_2$ blades whisking through the incident beam), and precisely 1/60 sec is
used for the actual parity violation measurement (determining the helicity
dependent transmission). A small phase slip is introduced so that the master
clock and the line frequency are again precisely in phase after 18 minutes.

   The LH$_2$ target (Fig.~7) has a flask of 0.10 m diameter and a length of
0.40 m. Special precautions have been taken to make the end windows of the
target flask optically flat and parallel. Maximum heat load of the target is
25 W with operation at approximately 5 W. By circulating the liquid hydrogen
rapidly (5 l/s ), density gradients are minimized. The target flask is 
remotely movable within $\pm $ 5 mm in two orthogonal directions at both ends to
position it on the `neutral axis'. The total scattering probability at 221.3
MeV by the 0.40 m long LH$_2$ is close to 4\%. The target flask length is
limited by multiple Coulomb scattering considerations; the various entrance
and exit windows of the LH$_2$ target (and all other energy degrading foils in
the beam) are kept to the minimally allowable number and thickness.

   The main detectors are two transverse electric field ionization chambers,
producing current signals due to direct ionization of the ultra-high purity
hydrogen gas by the beam. Field shaping electrodes plus guard rings ensure a
0.15 m wide by 0.15 m high by 0.60 m long sense region between the parallel
electrodes (negatively charged anode and signal plate), with the electric
field lines all parallel and perpendicular to the electrodes. The TRICS have
been designed for operation at -35 kV at one atmosphere; in practice they
are operated at a pressure of about 150 torr and a high voltage of -8 kV.
The entrance and exit windows are located at approximately 0.9 m from the
center of the TRICs to range out spallation products from proton interactions
with the stainless windows. (see Fig.~8 The design of the TRICs incorporated
considerations of noise due to $\delta$-ray production and due to recombination. 

   The proton beam energy in the downstream TRIC is on average 27 MeV lower
than in the upstream TRIC due to the difference in the  energy loss in the LH$_2$ target. Helicity
correlated energy modulations will cause a false A$_z$ due to the energy
dependence of the energy loss in the hydrogen gas of the TRICs. The
sensitivity to coherent energy modulations was determined using a RF
accelerating cavity placed upstream of IPM-1 in the beam line. The RF cavity
could produce coherent energy modulations with an amplitude of 600 eV in the
221.3 MeV proton beam. The measured sensitivity of (2.9 $\pm $ 0.3) 
$\times $ 10$^{-8}$
eV$^{-1}$ agrees very well with the prediction obtained in Monte Carlo
simulations of the experiment of $2.8 \times 10^{-8}$ eV$^{-1}$ and places a very
stringent constraint on the maximally allowable helicity correlated energy
modulation of the incident proton beam. Coherent energy modulations of the
extracted beam are caused by coherent modulations of the radial intensity
distribution at the cyclotron stripping foil; this converts position
modulations of the injected beam to energy modulations of the extracted beam.
The conversion factor has been estimated to be dE/dx$_{\rm 
injected}=$100 eV/$\mu$m which is in agreement with direct measurements 
made by applying large position modulations of the injected beam using 
electrostatic steering plates.  One source of the helicity correlated 
position modulations of the injected beam is helicity correlated 
energy modulations at OPPIS, which are converted to position 
modulations when the beam passes electrostatic steering plates on its 
way to injection into the cyclotron.  This process amplifies the 
primary coherent energy modulations at OPPIS by a factor of 
approximately 100, as determined in an auxiliary measurement using a 
magnetic spectrometer.  The sensitivity of $A_z $ to helicity 
correlated energy modulations at the ion source OPPIS was measured by 
applying a square wave voltage of 0.5 V amplitude to the sodium 
ionizer cell in OPPIS.  A sensitivity of $0.4 \times 10^{-8}$ per meV 
of coherent energy modulation at OPPIS was measured in agreement with 
the conversion factor measurement.  Consequently, coherent energy 
modulations at OPPIS are not to exceed about 1 meV.  The other source 
of helicity correlated position modulations of the injected beam are 
helicity correlated position modulations at OPPIS; the measured 
sensitivity of $A_z$ is approximately $0.1 \times 10^{-8}$ per nm of 
helicity correlated beam position modulation at OPPIS.  Consequently, 
coherent position modulations at OPPIS are not to exceed about 4 nm.  
Coherent energy and position modulations were monitored on a regular 
basis at OPPIS using a pair of electrostatic steering plates as a beam 
analyzer and an intensity profile monitor with 16 foil strips to 
measure the beam position downstream of the steering plates.  
Measuring on both sides of the OPPIS beam axis allowed for both energy 
and position modulations to be determined.  The precisions obtained 
after 20 minutes are $\pm$~2~nm in position and 0.2 meV in energy.  
But helicity correlated energy modulations cannot be directly measured 
at the parity violation measuring apparatus.  Therefore, an 
appropriate linear combination of the data taken with all possible 
helicity combinations and spin precessions is being used to remove any 
false $A_z $ due to helicity correlated energy modulations.

   To date a series of data taking runs have taken place. Of these the second
and fourth data taking runs have been analyzed. The preliminary result
is shown in Fig.~9 and compared with the theoretical prediction of Driscoll and
Miller$^{33}$.  The theoretical
prediction of Driscoll and Miller is based on the Bonn potential to represent the strong interactions
together with the weak meson-nucleon coupling constants as given by
Desplanques, Donoghue, and Holstein$^1$. It treats Coulomb effects and
relativistic distortions, but inelastic effects are treated in an indirect
manner. Even though the Driscoll and Miller theoretical prediction
overestimates the size of $A_z $ at the lower energies (correctible through
adjustment of the weak $\rho$- and $\omega$-nucleon coupling constants), it
exhibits the expected energy behavior. 
Figure 10 compares various theoretical predictions with the low 
energy $p-p$ parity violation data.  The theoretical prediction of 
Grach and Shmatikov$^{37}$ is based on a quark model which emphasizes the 
second transition ${^3P_2}$-${^1D_2}$ contribution leading to 
a rather different energy dependence.  The theoretical predicton of 
Iqbal and Niskanen$^{32}$ has a $\Delta$ isobar contribution included.  
The theoretical
prediction of Driscoll and Meissner is based on a self consistent calculation, with both
weak and strong vertex functions obtained with a chiral soliton model.
Note the small value of $A_z $ predicted at 221 MeV. Figure~11 gives the
constraints placed on the weak meson-nucleon coupling constants by the $p-p$
parity violation experiments. It is anticipated that continued data taking
and data analysis will give the 221~MeV experimental result uncertainties less than
 $2 \times 10^{-8}$ in statistics and in systematics. 
 
\section{Proton-proton Parity Violation Measurements at Higher 
Energies}

   A further $p-p$ parity violation experiment is in preparation at TRIUMF
at an energy of 450 MeV.$^{38}$ With an inelasticity of at most 10\% of the total
cross section, the expected impact on the deduction of the combination of
weak meson-nucleon coupling constants $h^{pp}_\rho$ and $h^{pp}_\omega$ is
considerably less than the aimed-for experimental errors of $2 \times 10^{-8}$ in
statistics and $2 \times 10^{-8}$ in systematics. The measurement can be made with
minimal changes to the apparatus of the present TRIUMF $p-p$ parity violation
experiment, except for modifications to the PPMs to allow higher count rates.
The combination of measurements at 221 MeV and 450 MeV would give an
independent determination of the weak meson-nucleon coupling constants
$h^{pp}_\rho$ and $h^{pp}_\omega$.

   Measurements of $A_z$ in $p-p$ scattering have also been proposed at COSY
of the Forschungszentrum J\"ulich as a fixed target experiment near 230 MeV and
at a higher energy as a cooler ring experiment.$^{39}$ The choice of the latter
energy (or better an energy close to the maximum energy of COSY) is in part
motivated by the earlier 5.13 GeV measurement of $A_z$ (on a water target)
at the ZGS of Argonne National Laboratory, which resulted in 
$A_z$ = $(26.5 \pm 6.0 \pm 3.6) \times 10^{-7}$.$^{40}$ This result is an order of
magnitude larger than what is expected using conventional scaling arguments.
It must be remarked that various re-evaluations of the experiment have not
come across any flaw (in the way the experiment was conducted) that could have
led to such a large false  $A_{z}$. It has also been pointed out that when
Glauber shadowing is taken into account, the $p-p$ parity violating $A_z$
increases by as much as 40\%.$^{41}$ Figure 12 shows the energy dependence of $A_z$;
note the logarithmic scale used for the abscissa.   

   The theoretical curve$^{42}$ that matches the experimental data at 800 MeV
and at 5.13 GeV has been normalized to the 5.13 GeV datum. This calculation is
based on a diquark model introducing a parity violating component in the
nucleon wave function. The authors find an important role for diagrams
in which the weak interactions between the members of a vector diquark in the
polarized proton is accompanied by the strong interaction between that diquark
and a quark of the other nucleon. The theoretical curve exhibits a steep
increase for $A_z$ with increasing energy. It predicts a value for $A_z$ at 20 GeV
of the order 10$^{-5}$. Note that a 200 GeV experiment has placed an upper limit
(95\% C.L.) on the $p-p$ longitudinal analyzing power $A_z$ of $5.7 
\times 10^{-5}$ (Ref.~43). 
The theoretical interpretation$^{42}$ of the unexpectedly large result for $A_z$
at 5.13 GeV has created a great deal of controversy$^{44}$. Many of the earlier
theoretical predictions for $A_z$ give values of the order of $10^{-7}$ at
5.13 GeV; see for instance Ref.~45. Clearly, the 5.13 GeV result presents a
great challenge both in obtaining its confirmation through a new 5 GeV
experiment and in obtaining a self-consistent theoretical explanation.

   If confirmed experimentally, the need for a further experiment at an energy
of tens of GeV becomes well established indeed. Such a second higher energy
measurement of parity violation in $p-p$ scattering could be made at the AGS
of Brookhaven National Laboratory$^{46}$ or at the proposed Japanese Hadron
Facility. It has been pointed out that a storage ring/internal target
environment would allow for innovative methods in performing such an
experiment, quite different from the methods used in all previous $p-p$ and
proton-nucleus parity violation experiments$^{47}$. A schematic illustration
of a possible $p-p$ parity violation experiment in a storage ring with an
internal target is shown in Fig.~13. The polygonal ring has a twelvefold
symmetry. The Siberian snake of the ``first kind" precesses the spin direction
of the polarized protons of the beam by 180 degrees about the longitudinal
axis. When energized the Siberian snake causes the stable spin orientation
in each straight section of the ring to be horizontal, rather than vertical.
Choosing the appropriate beam energy, corresponding to half integer spin tune
of G$\gamma $ = 7.5, or 13.5, etc. (where G is the anomalous magnetic moment
of the proton and $\gamma$ is the standard relativistic factor) one arrives at
a stable spin direction for the various straight sections of the ring as
indicated (beam energies of 2.99 GeV, 6.13 GeV, etc.). The attenuation of the
stored beam has to be dominated by the nuclear scattering and not by Coulomb
scattering. Consequently an energy of a few GeV may be too low for a storage
ring parity violation experiment. One should note that the stable spin
direction for any bombarding energy is longitudinal in the straight section
located diametrically opposite the Siberian snake (see Fig.~13). Consequently,
this is the location to perform the transmission experiment with an internal
gaseous high purity hydrogen target. The unwanted transverse polarization
components of the beam at this location reverse sign upon each successive
passage of the Siberian snake and are therefore very effectively cancelled.
As stated above for the 221 MeV TRIUMF $p-p$ parity violation experiment, these
transverse polarization components and their first moments can cause
significant systematic errors in external, fixed target experiments. The
spin direction of the stored beam can be flipped rapidly by an `adiabatic
fast passage' technique using an `RF field' also indicated in Fig.~13 and
independently at a lower rate by helicity reversal at the polarized ion source
between successive beam injection into the ring$^{47}$. The attenuation
experiment consists of determining the helicity dependence of the stored beam
lifetime by a current transducer. The resolution of the current transducer is
one of the determining factors for the precision attainable in a storage ring
experiment. Such current monitors are being developed$^{48}$. The use of a
windowless gaseous pure hydrogen target as well as the cancellation of the
extraneous transverse polarization components are the distinct advantages
of the stored beam/internal target environment. The ring will also act as
a magnetic spectrometer, so that the experiment becomes insensitive to a
false parity violation signal caused by weak decays of the hyperons produced
in $p-p$ interactions. Of concern are, however, the interactions of the stored
beam with the residual gas in other sections of the ring, where the
polarization direction of the beam is not longitudinal, or the presence of
beam halo interacting with the ring structure, as well as helicity correlated
changes in beam emittance in successive injection cycles or upon `RF field'
induced spin flips. Clearly, a thorough evaluation through simulations is
needed to underpin the asserted advantages of the storage ring/internal 
target environment for any $p-p$ parity violation experiment.
 
   Information about the flavor non-conserving hadronic weak interaction stems
from non-mesonic decays of hypernuclei. Increasingly for heavier and heavier
hypernuclei, the $\Lambda$ inside the nuclear medium does not decay through
the (by the Pauli principle blocked) mesonic channel, but through the
 $\Delta$S = 1 non-mesonic channel, $\Lambda-N \to N-N$. Less theoretical
encumbrance is provided by studying the inverse reaction $ p-n \to p-\Lambda $ 
(threshold energy 368.5 MeV). However, the expected small cross section of
less than $10^{-13}$ times the $ p-n$ elastic scattering cross section poses a
great experimental challenge, which has not yet been met. Most promising,
possibly, appears a study of quasi-free $\Lambda$ production
in $ d-p \to p-p-\Lambda $ below the threshold for associated production
 $ d-p \to p-n-\Lambda-K^+$. Detection of the two protons allows for a
determination of the missing mass (selecting the mass of the $\Lambda$
permits suppression of $p-p-n$ events), while the observation of the decay of
the $\Lambda$ into $p-\pi^-$ excludes strong interaction $\pi^-$ production
(like $ p-p-p-\pi^-$ events). Delayed $\Lambda $ decay occurs on average 50~mm
downstream of the $\Lambda $ production vertex. Complete reconstruction of the
 $\Lambda \to p-\pi^-$ decay presents a highly desirable additional kinematical
constraint on the weak $\Lambda $ production. A study of the reaction
$ d-p \to p-p-\Lambda $ has been proposed at COSY$^{49}$; the very small cross
section poses great demands on the quality and intensity of the longitudinally
polarized beam and on the detection system with second and possibly third
level fast triggers. A study of the reaction $ p-n \to p-\Lambda $ on a nuclear
target has been proposed at RCNP$^{50}$. Again the proposed detection system is
quite complex and relies on detection of the $\Lambda $ decay products in
a region shielded from direct view by the target. Recent theoretical
predictions of the cross sections and analyzing powers and their energy
dependence are given in Ref. 51,52. The observables are found to be rather
sensitive to the opening of the $\Sigma^0$ production channel (at 540.0 MeV
for $ p-n \to  p-\Sigma^0$).

\section{Conclusions}

   Several new $N-N$ parity violation experiments are currently being executed;
these allow one to anticipate a determination in the near future of
the weak meson-nucleon coupling constants and a resolution of the present
uncertainty about the weak pion-nucleon coupling constant. The TRIUMF
$p-p $ parity violation experiment has produced a first result and is in the
final stages of data taking. A measurement of the parity violating neutron
spin rotation in helium is ongoing at NIST, while a measurement of the parity
violating asymmetry in the capture of longitudinally polarized neutrons by
hydrogen has started at LANSCE.

\newpage

\renewcommand{\arraystretch}{2}
\begin{sidewaystable}
Table I  Weak meson-nucleon couplings constants\\
\noindent
\begin{tabular*}{20cm}{@{\extracolsep{\fill}}cr@{}c@{}lccr@{}c@{}lccc|cr@{}c@{}l}
Coupling & \multicolumn{10}{c}{Theoretical} & \multicolumn{5}{c}{Experimental}
\\ 
\cline{1-1}\cline{2-12} \cline{13-16}
     & \multicolumn{3}{c}{range}  & `best value' & value & \multicolumn{3}{c}{range}  & `best value' & value     
     & value      & best fit & \multicolumn{3}{c}{range} \\
       & \multicolumn{3}{c}{{\sc (ddh)}} & {\sc (ddh)}  & {\sc (dz)} & \multicolumn{3}{c}{{\sc (fcdh)}} & {\sc (fcdh)}
        & {\sc (d)} & {\sc (km)} &         & \\
\hline
$f_\pi$    & 0     & $\to$ & 11.4 & 4.6   &  1.1  & 0     & $\to$ & 6.5  &  2.7 &  2.7 &  0.19 &  2.3& 0 &$\to$& 11  \\
$h^0_\rho$   & -31   & $\to$ & 11.4 & -11.4 & -8.4 & -31   & $\to$ & 11   & -3.8 & -6.1 & -1.9  & -5.7 &-31 &$\to$& 11  \\
$h^1_\rho$   & -0.38 & $\to$ & 0    & -0.19 &  0.38 & -1.1  & $\to$ & 0.4  & -0.4 & -0.4 & -0.02 & -0.2&-0.4 &$\to$& 0.0 \\
$h^2_\rho$   & -11.0 & $\to$ & -7.6 & -9.5  & -6.8 & -9.5  & $\to$ & -6.1 & -6.8 & -6.8 & -3.8  & -7.6 &-11 &$\to$&-7.6 \\
$h^0_\omega$ & -10.3 & $\to$ &  5.7  & -1.9 & -3.8 & -10.6 & $\to$ & 2.7  & -4.9 & -6.5 & -3.8  & -4.9 &-10 &$\to$& 5.7 \\
$h^1_\omega$ & -1.9  & $\to$ & -0.8 & -1.1  & -2.2 & -3.8  & $\to$ & -1.1 & -2.3 & -2.3 & -1.0  & -0.6 &-1.9 &$\to$& -0.8 \\
\hline \hline   
\end{tabular*}
\end{sidewaystable}

\begin{sidewaystable}
\noindent
\parbox{16cm}{
Table II  Comparison of calculated and measured sensitivities to systematic
          errors and the precision required in measuring the various
          parameters.}
\begin{tabular*}{20cm}{@{\extracolsep{\fill}}p{4cm}p{4cm}p{4cm}p{4cm}}
\hline \hline
Beam Parameter & Sensitivity (calculated) & Sensitivity (measured) & Measure to \\
\hline
$<xP_y>$              & +3 $\times$ $10^{-5}$ mm$^{-1}$  & +5 $\times$ 10$^{-5}$ mm $^{-1}$  & 12 $\times$ 10$^{-5}$ mm  \\  
$<yP_{x}>$            & -3 $\times$ $10^{-5}$ mm$^{-1}$  & -8 $\times$ 
10$^{-5} $ mm$^{-1}$  & 8 $\times$ 10$^{-5}$ mm\\
Position Modulation  ($<x> \Delta x)$
  & 3.2 $\times$ 10 $^{-5}$ mm$^{-2}$ & 1.2 $\times$ 10$^{-4}$ mm$^{-2}$ 
   & 5 $\times$ 10 $^{-5}$ mm$^{2}$\\ 
Size modulation       & 6.5 $\times$ $10^{-5}$ mm$^{-2}$  & 2.9$\times$ 
10$^{-4} $ mm$^{-2}$  & 3 $\times$ 10$^{-5}$ mm$^{2}$\\
\hline
Energy Modulation & 1.4 $\times$ 10$^{-8}$/eV & (1.5 $\pm$ 0.2) $\times$ 
10 $^{-8}$/eV &  0.4 eV\\
                  \hline
Current Modulation & 3 $\times$ 10$^{-9}$       & 5 $\times$ 10$^{-9}$ &  1.0 $\times$ 10$^{-5}$ \\
                   & (for $\Delta I/I = 10^{-5})$ & (for $\Delta I/I \times 
                   10^{-5}$)  &\\
                   \hline \hline 
                   \end{tabular*}
\end{sidewaystable}

\clearpage


\begin{center}
\epsfig{figure=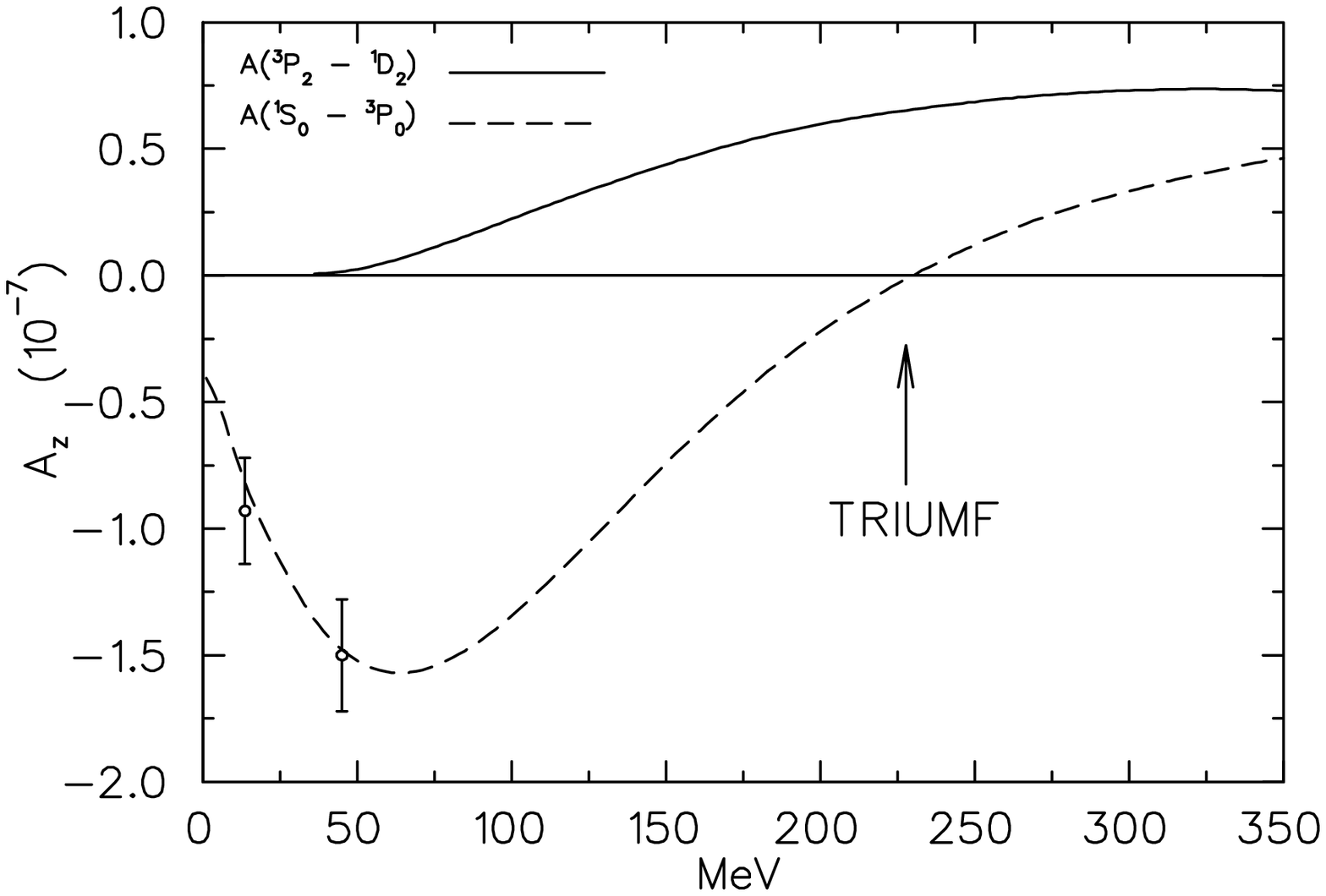,width=\linewidth}
\end{center}

\noindent
 Fig.~1.  Contributions to $A_z$ of the first two parity violating transitions in
        $p-p$ scattering as function of energy.


\begin{center}
\epsfig{figure=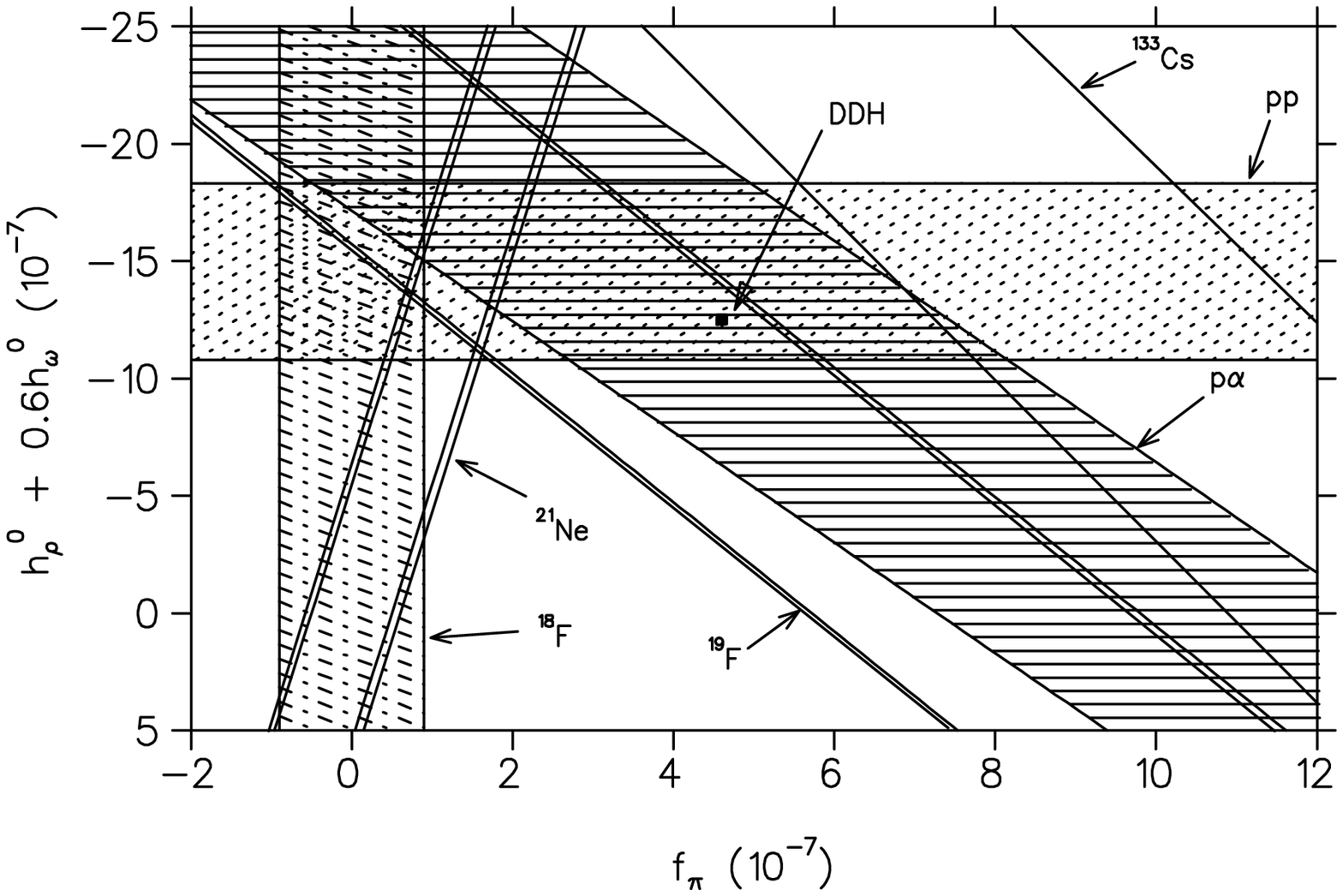,width=\linewidth}
\end{center}
\noindent
 Fig.~2.  Plot of the constraints on the isoscalar and isovector weak meson-
        nucleon coupling constants. 
        
\begin{center}
\epsfig{figure=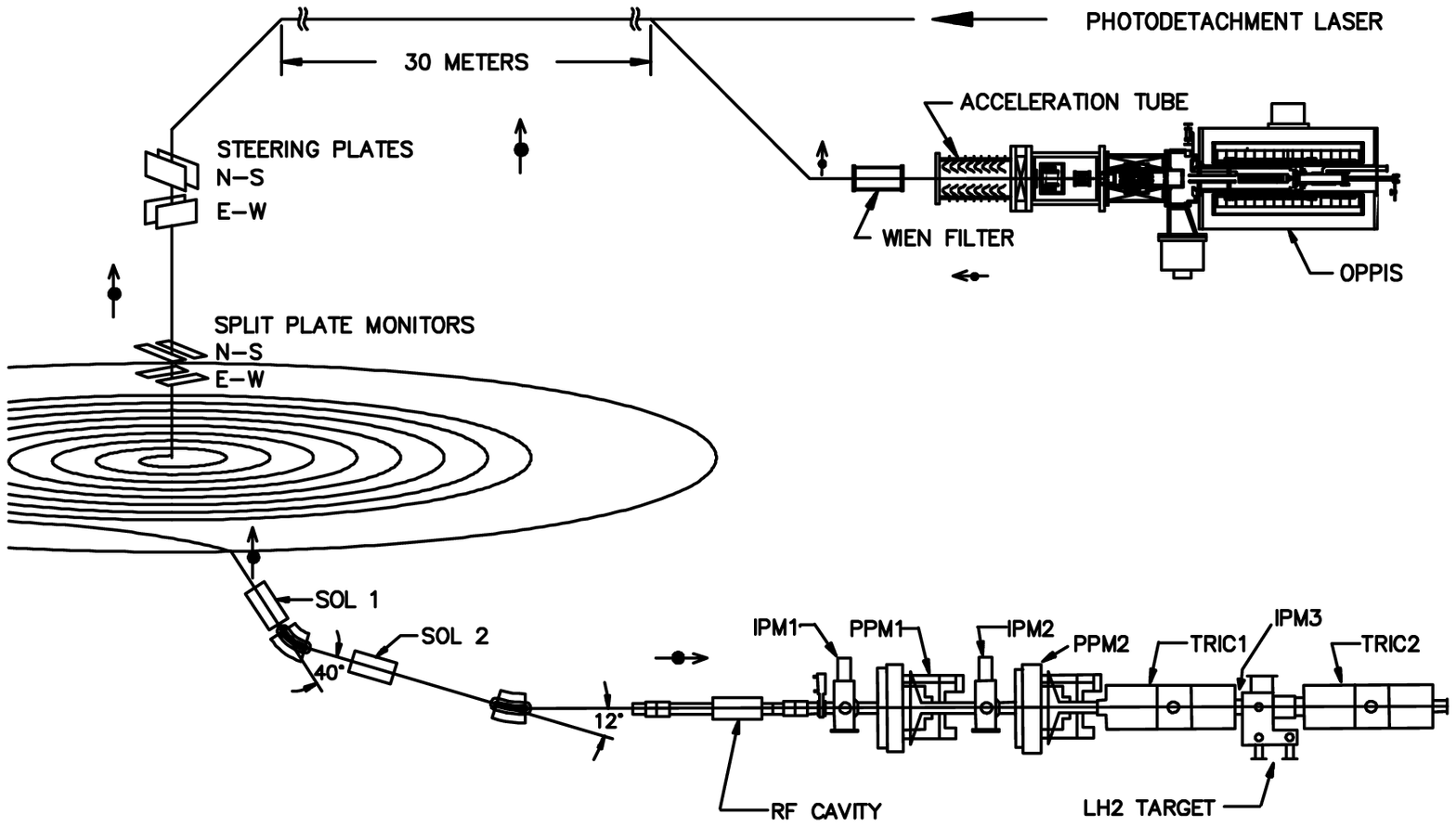,width=\linewidth}
\end{center}
\noindent
 Fig.~3.  General layout of the TRIUMF $p-p$ parity violation experiment. (OPPIS:
        Optically Pumped Polarized Ion Source; SOL: spin precession solenoid
        magnet; IPM: Intensity Profile Monitor; PPM: Polarization Profile
        Monitor; TRIC: Transverse Electric field Ionization Chamber).  One
        of eight possible spin directions is indicated.

\begin{center}
\epsfig{figure=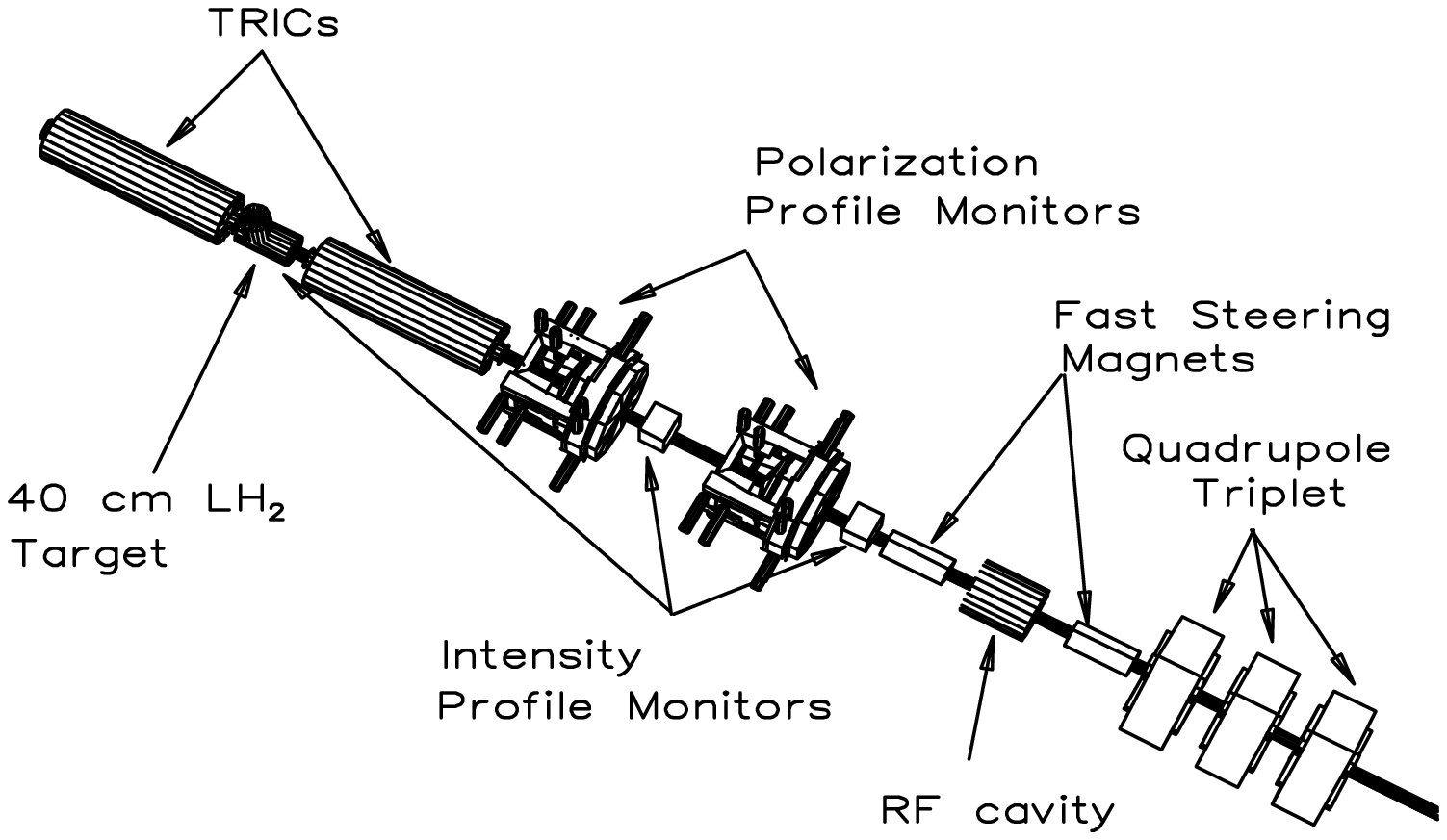,width=\linewidth}
\end{center}
\noindent
 Fig.~4.  Three dimensional view of the TRIUMF $p-p$ parity violation detection
        apparatus. (note the beam entering from the lower right)

\begin{center}
\epsfig{figure=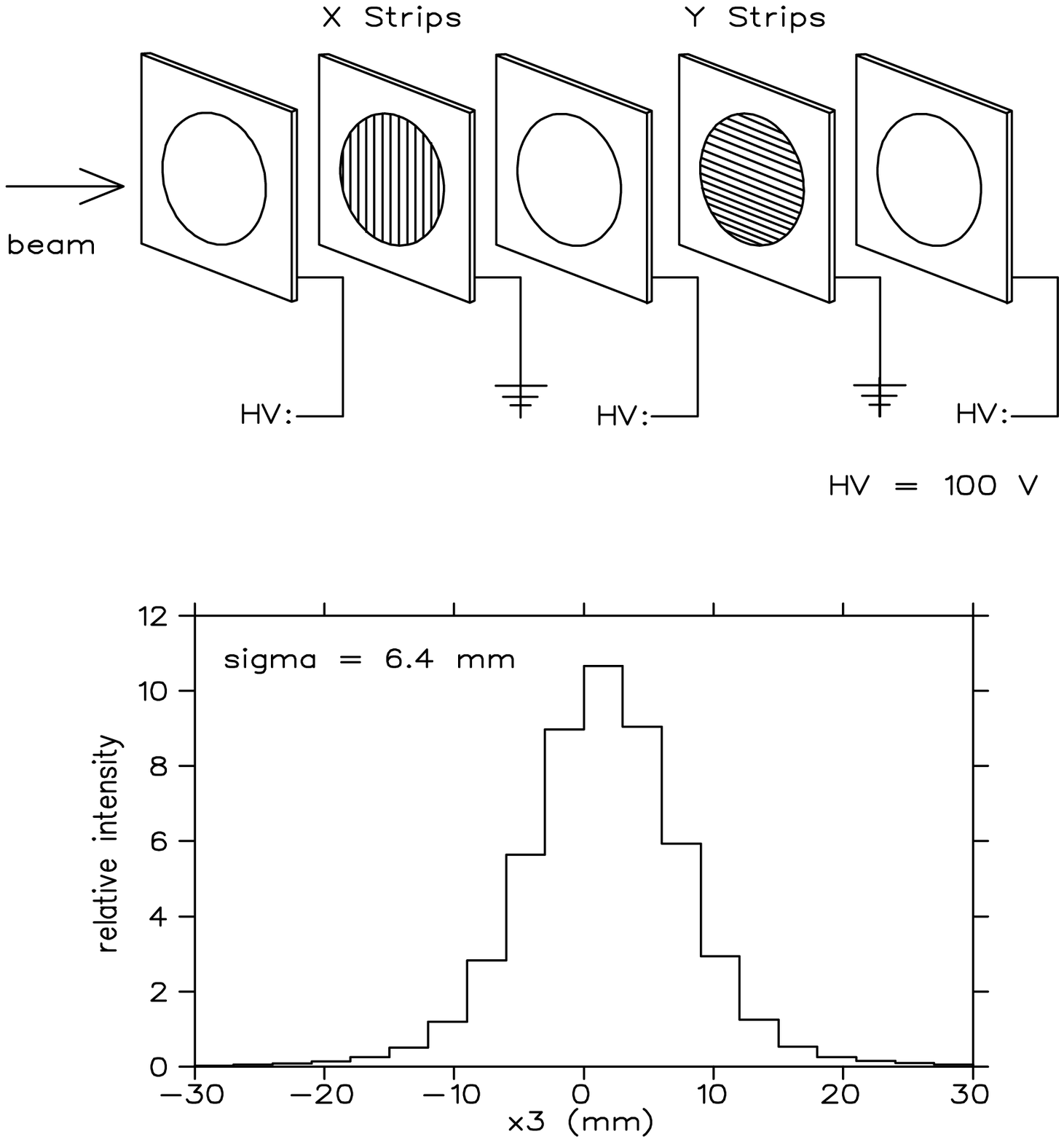,width=\linewidth}
\end{center}

\noindent
 Fig.~5.  a) Schematic representation of the assembly of foil harps of each IPM.
           The foil harps are mounted on G10 frames with apertures of 76~mm
           (for IPM-3 140~mm).
        b) Typical horizontal beam intensity profile produced by a harp of 31
           aluminium strips (3.00 mm center to center) of IPM-3. The intensity
           profile has a $\sigma_x$ of 6.4 mm.   

\begin{center}
\epsfig{figure=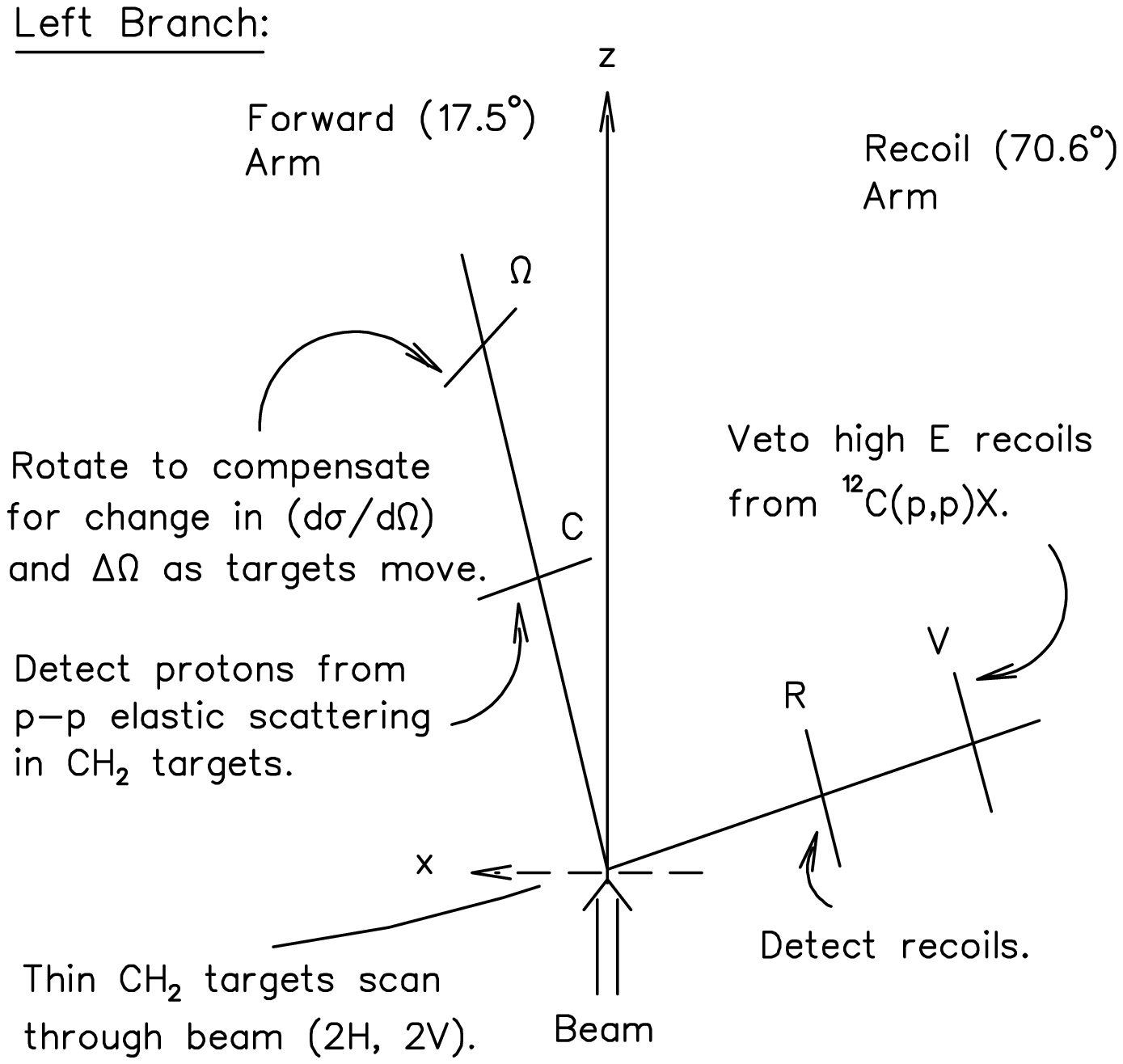,width=\linewidth}
\end{center}

\noindent
 Fig.~6.  Schematic representation of one of the four detector assemblies of
        each PPM.

\begin{center}
\epsfig{figure=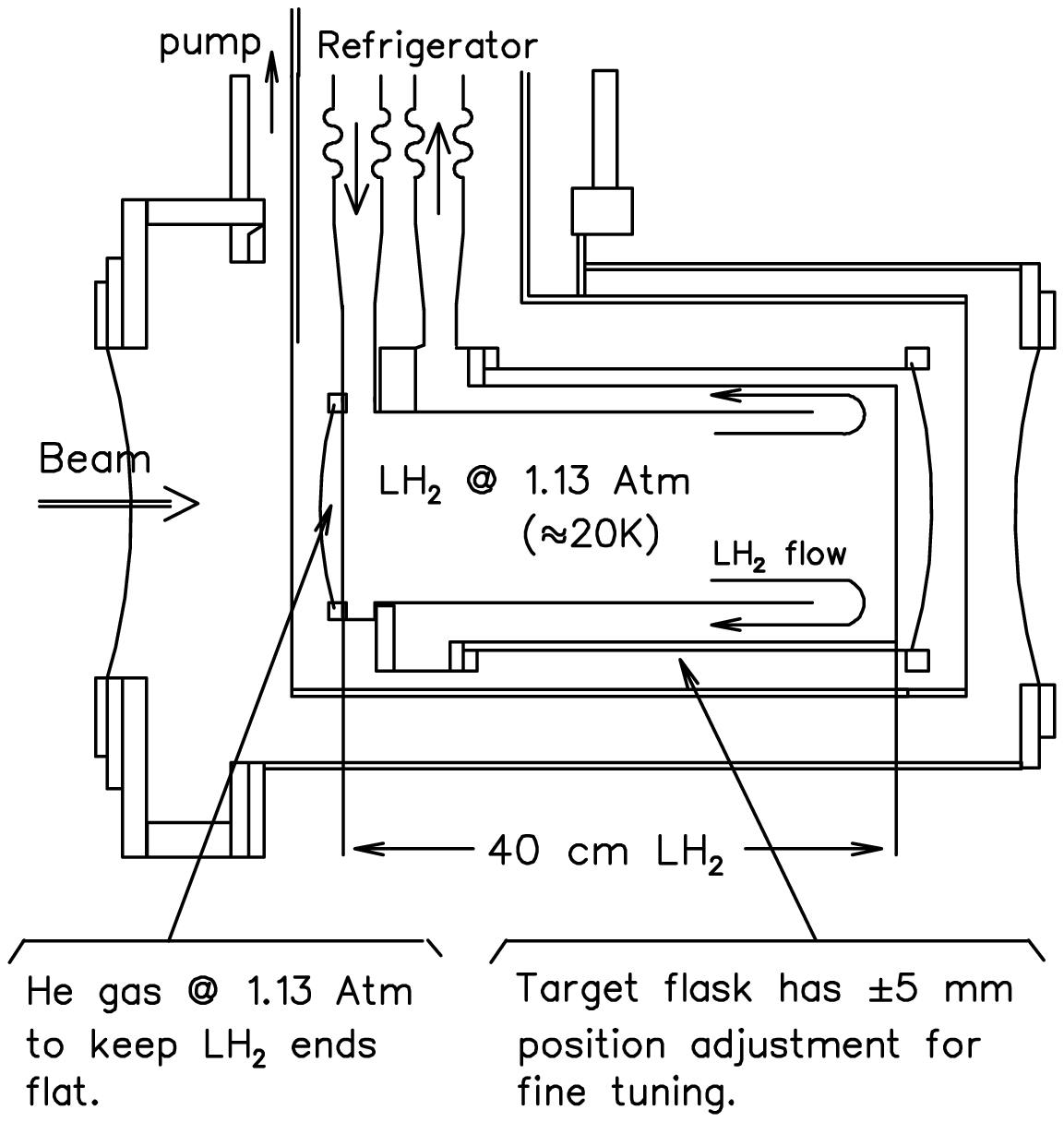,width=\linewidth}
\end{center}

\noindent
 Fig.~7.  Engineering drawing of the LH$_2$ target.

\begin{center}
\epsfig{figure=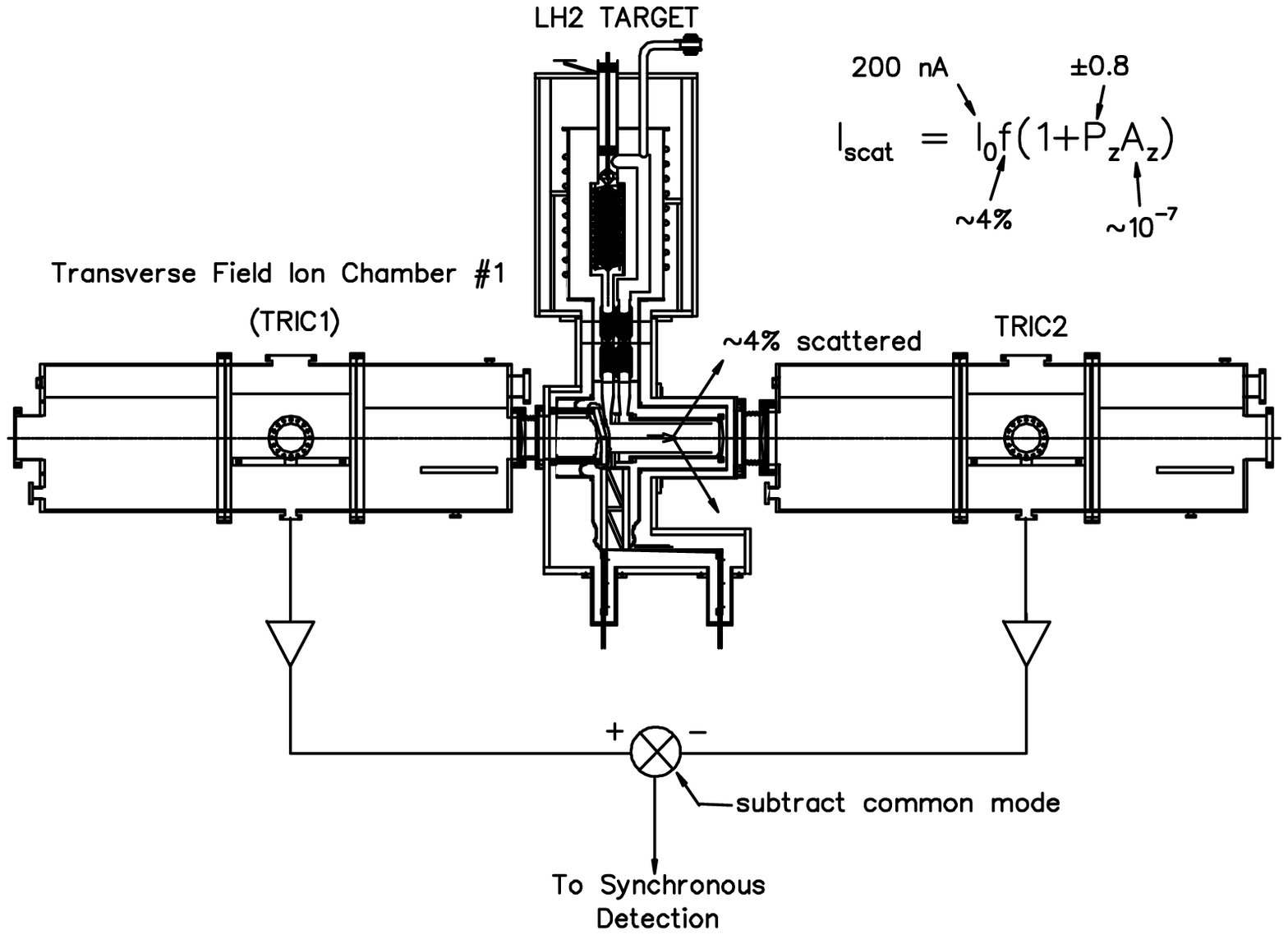,width=\linewidth}
\end{center}

\noindent
 Fig.~8.  Parity violation data taking apparatus: TRIC1 and TRIC2 are the
        transverse electric field ionization chambers with a 0.60 m long by
        0.15 m wide by 0.15 m high sense region and contain ultrapure hydrogen
        gas at approximately 150 torr pressure; the LH$_2$ target is 0.40 m long. 

\begin{center}
\epsfig{figure=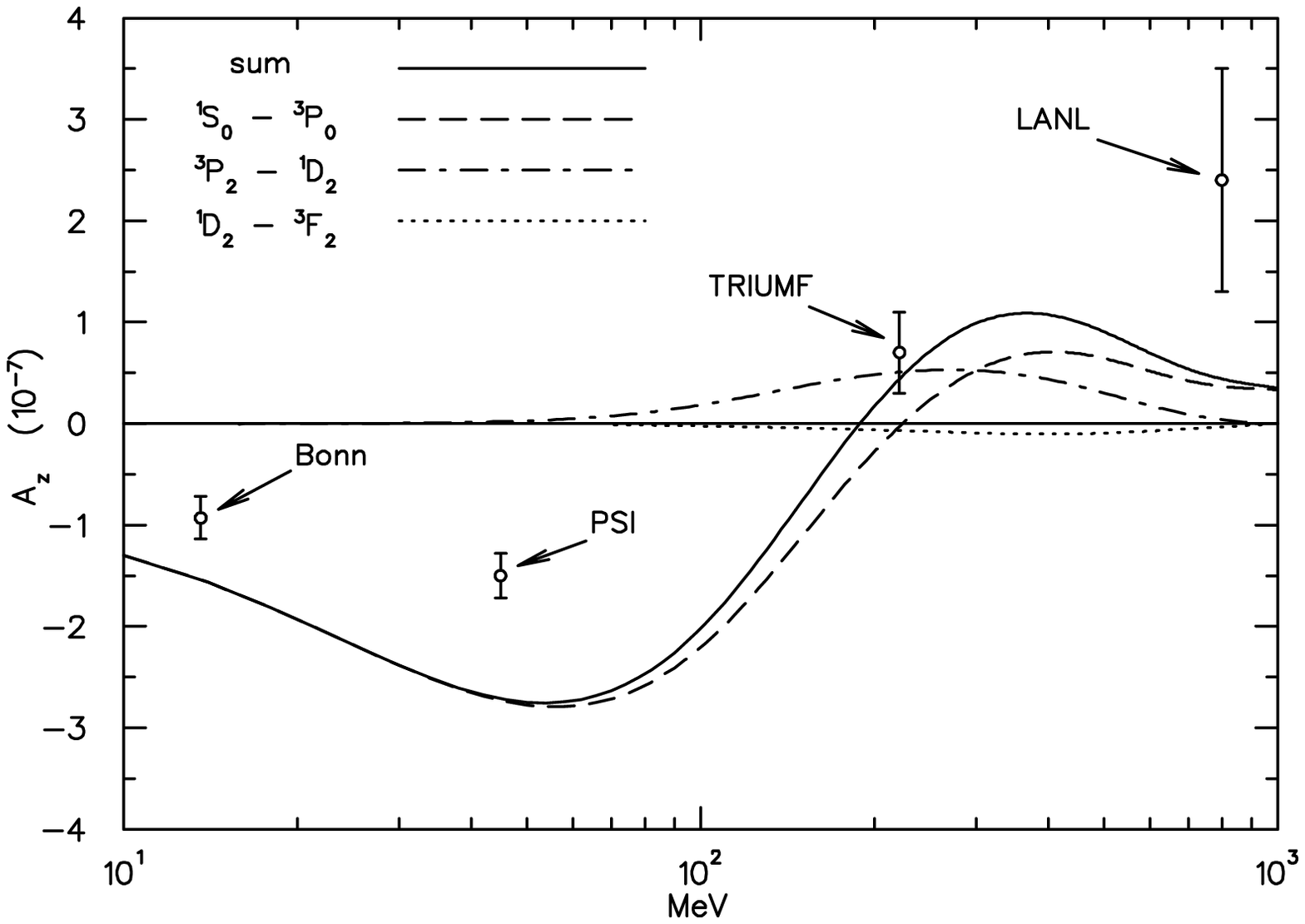,width=\linewidth}
\end{center}

\noindent
 Fig.~9. Energy dependence of the $p-p$ parity violating longitudinal analyzing
        power $A_z$. The curves give the total and the individual contributions
        of the first three parity violating transition amplitudes as
        calculated by Driscoll and Miller [Ref.33] in a weak meson exchange
        model. 
        Note the logarithmic energy dependence of the abscissa. The 221~MeV
        datum is
        a partial result of the TRIUMF experiment.


\begin{center}
\epsfig{figure=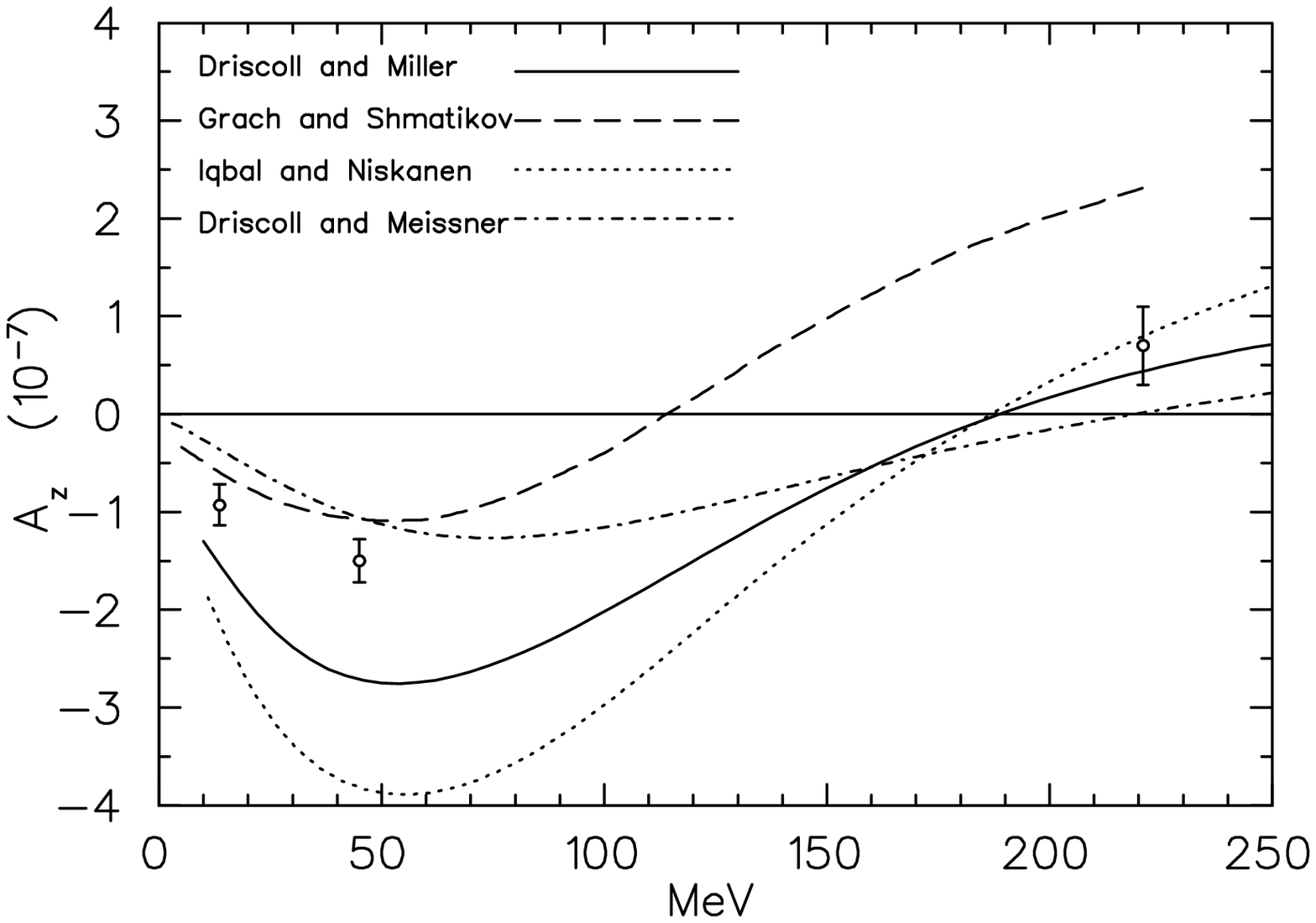,width=\linewidth}
\end{center}
\noindent
 Fig.~10. Theoretical predictions by Driscoll and Miller [Ref. 33],  
 Grach and Shmatikov [Ref. 37], Iqbal and Niskanen [Rev. 32], and 
 Driscoll and Meissner [Ref. 34] and the low energy $p-p$ parity 
 violating longitudinal analyzing power ($A_z$) data. 
  

\begin{center}
\epsfig{figure=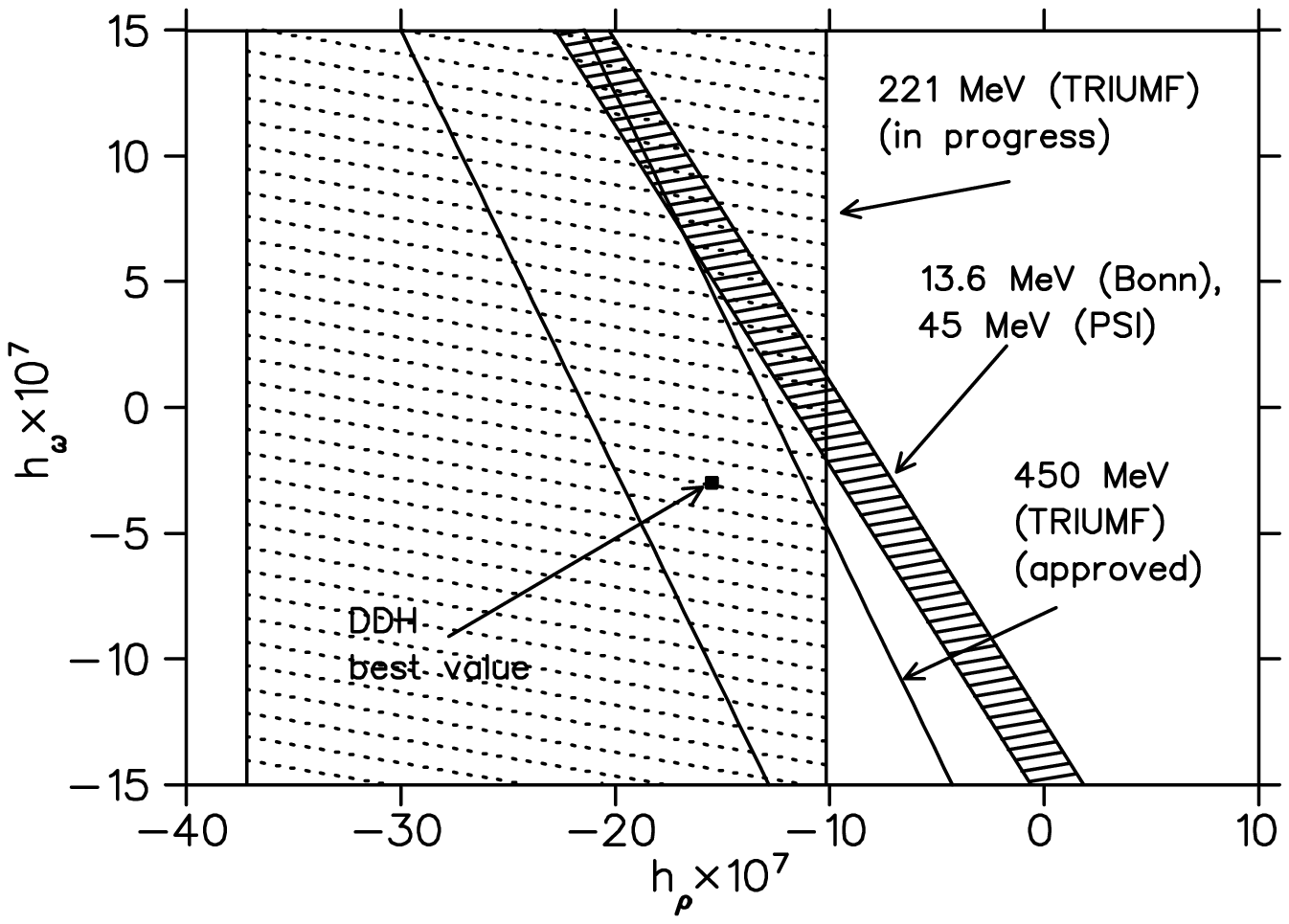,width=\linewidth}
\end{center}
\noindent
 Fig.~11.  Plot of the constraints placed on the weak meson-nucleon coupling
        constants by the $p-p$ parity violation experiments. The limit placed
        by the low-energy measurements is shaded; the DDH "best value"
        is indicated.  The error of a 450 MeV measurement is assumed to be
        $\pm 2 \times 10^{-8}$.


\begin{center}
\epsfig{figure=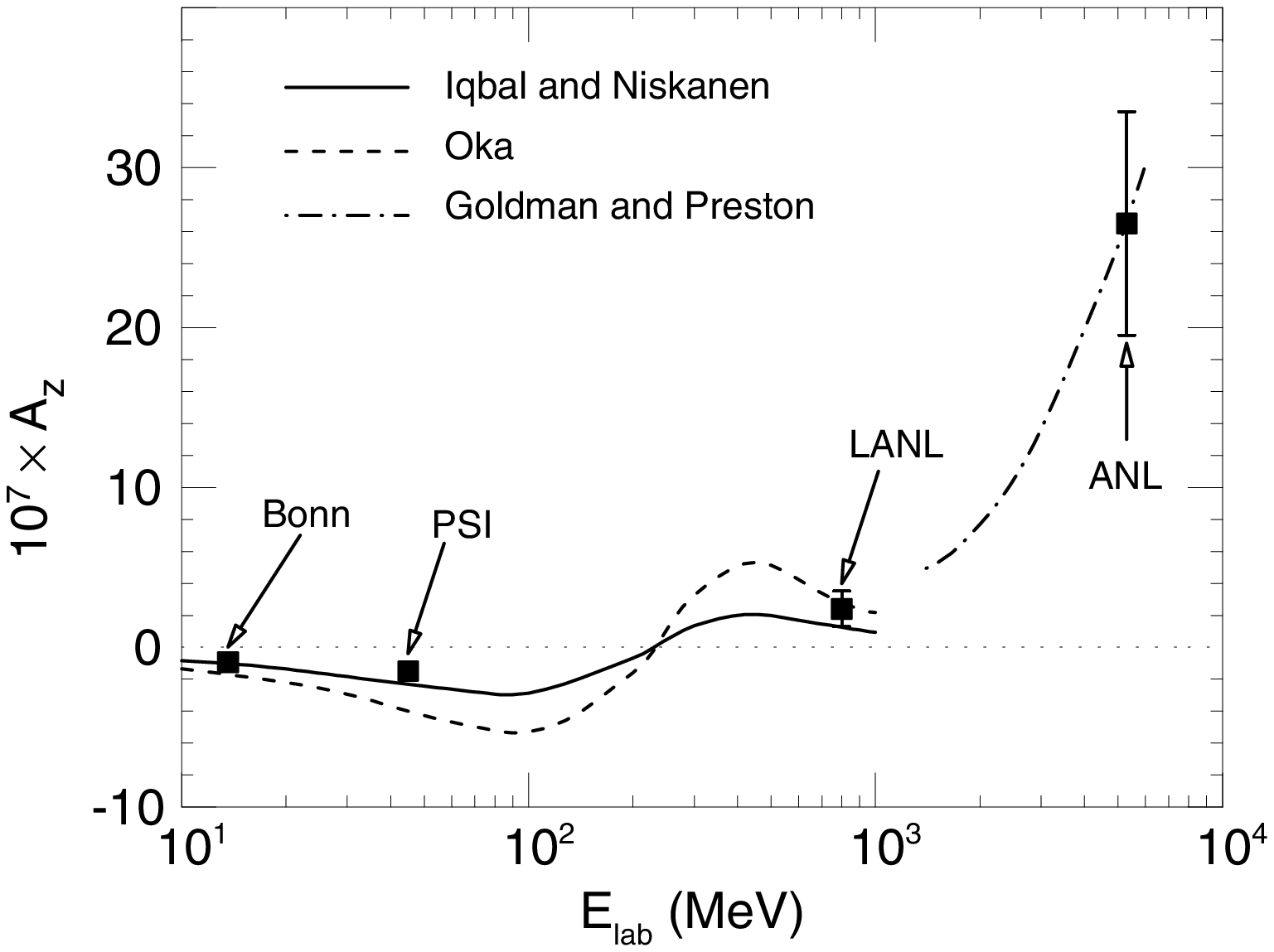,width=\linewidth}
\end{center}
 \noindent
  Fig.~12. Energy dependence of the $p-p$ parity violating longitudinal analyzing
        power in the energy range 10 MeV to 10 GeV. The solid and dashed
        curves are from Iqbal and Niskanen [Ref.~32] and Oka [Ref.~45], 
        respectively, based on weak meson exchange models; the dot-dashed
        curve is from Goldman and Preston [Ref.~42] and is described in the text.
 

\begin{center}
\epsfig{figure=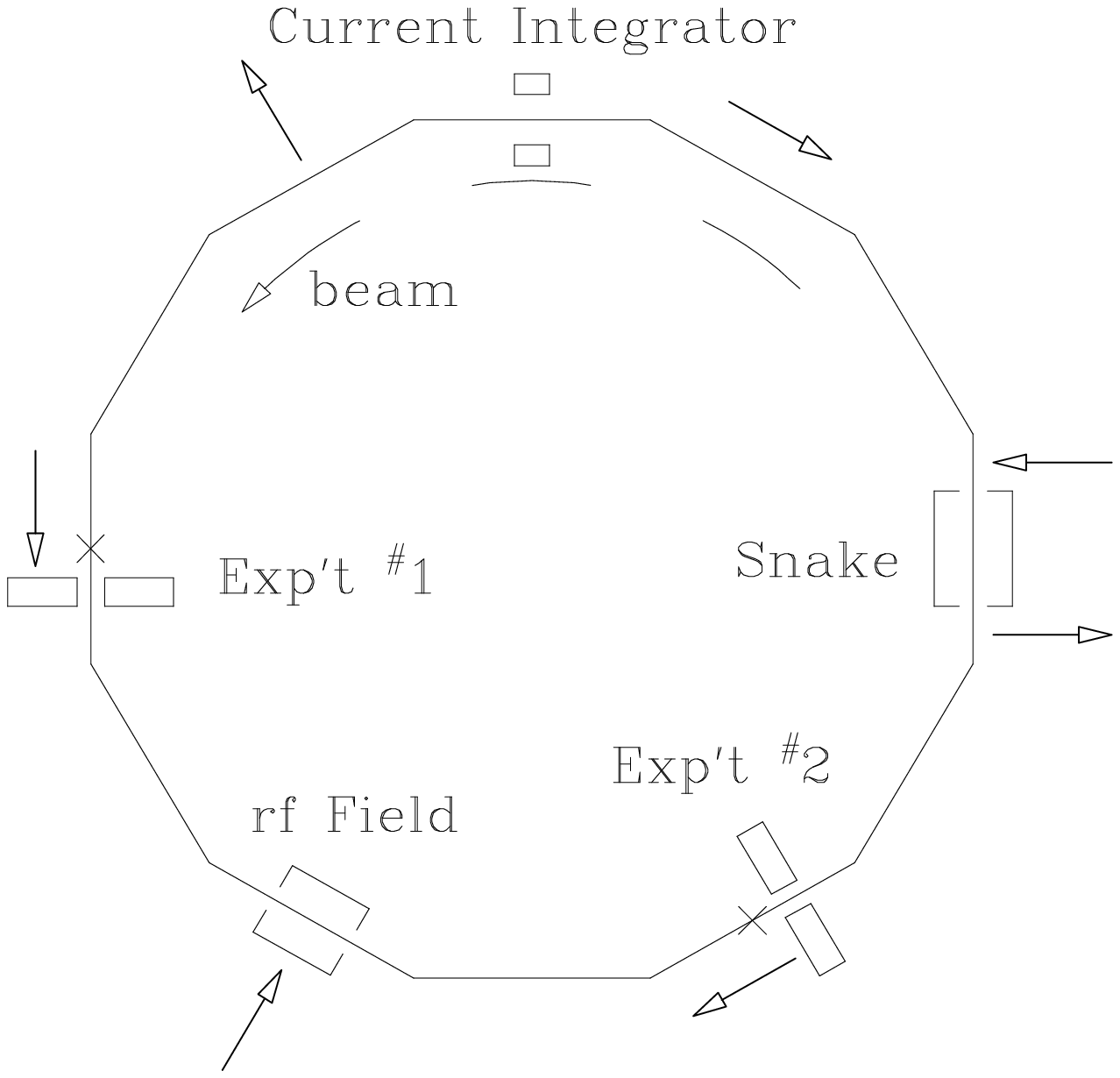,width=\linewidth}
\end{center}
 \noindent
 Fig.~13. Schematic layout of a possible $p-p$ parity violation experiment in a
        storage ring with internal targets. The target for a transmission
        (attenuation) experiment would be mounted at the location labeled
        `Exp't \#1'. Note the location of the Siberian snake of the first
        kind (figure is from reference 47).

\end{document}